# Short Term Stress of COVID-19 on World Major Stock Indices

**Muhammad Rehan** (  rehan.muhammad2626@hotmail.com )

   Gaziosmanpasa University    https://orcid.org/0000-0001-5056-5307

**Jahanzaib Alvi**

  Iqra University    https://orcid.org/0000-0001-9145-6545

**Süleyman Serdar KARACA,**

   Gaziosmanpasa University

**Research Article**

**DOI:** https://doi.org/10.21203/rs.3.rs-49510/v1

# Abstract

The main objective of this study is to check short term stress of COVID-19 on the American, European, Asian, and Pacific stock market indices, furthermore, the correlation between all the stock markets during the pandemic. Secondary data of 41 stock exchange from 32 countries have been collected from investing.com website from 1$^{st}$ July 2019 to 14$^{th}$ May 2020 for the stock market and the COVID-19 data has been collected according to the first cases reported in the country, stocks market are classified either developed or emerging economy, further divided according to the subcontinent i.e. America, Europe, and Pacific/Asia, the main focus in the data is the report of first COVID-19 cases. The study reveals that there is volatility in the all the 41 stock market (American, Europe, Asia, and Pacific) after reporting of the first case and volatility increase with the increase of COVID-19 cases, moreover, there is a significant negative relationship between the number of COVID-19 cases and 41 major stock indices of American, Europe, Asia and Pacific, European subcontinent market found more effected from the COVID-19 than another subcontinent, there is Clustering effect of COVID-19 on all the stock market except American's stock market due to smart capital investing.

# 1. Introduction

Coronavirus is known as COVID-19 which affects the Wuhan, China in December 2019 and became the cause of the crisis in Hubei China and then for the rest of the world, all of sudden COVID-19 became a global pandemic. The entire world faced volatility in the stock market and a significant decline in the equity market. This is the biggest volatility level seen in the United State stock market after October 1987 and December 2008. (Baker, S., Bloom, N., Davis, S. J., Kost, K., Sammon, M., & Viratyosin, T. 2020)

The WHO (World Health Organization) officially declared a global pandemic to the COVID-19 outbreak on 11$^{th}$ March 2020. According to the WHO number of confirmed cases has been reached up to 4 307 287 on 14$^{th}$ May 2020 and it continues increasing day by day, Further as per reported by the WHO 216 countries has been affected by this pandemic and most number of confirmed cases has been reported in the USA with 1,361,522 confirmed cases (WHO). The COVID-19 has significant effect the world economy in short term as well as in the long term, in shorter-term consequence is limited activity in the economy due to strict lockdown, in the longer-term impact of the COVID-19 many small businesses will be closed and unemployment will be increased, the number of industries will suffer i.e. tourism airlines and hotels (Zhang, D., Hu, M., & Ji, Q. 2020).

The question of this research is addressed the predominant importance equally for the market policymaker, institutional investors, and individual investors. 12% decrease has been recorded in the Dow Jones Industrial Average index which was the 2$^{nd}$ highest decline in the market history since 124 years on 16$^{th}$ of March 2020, although US government has taken many actions intended to improve the market which includes economy relief program and fiscal stimulus package the market did not improve (Gormsen & Koijen 2020).

This paper aims to study the short term stress of COVID-19 on the performance of the major stock indices of the 32 countries. In this research to recognize the systemic risk pattern in the stock market we are going to answer the following questions through available data, How will we react to the stock market during this pandemic? Does systemic risk escalate all over the world? Do the clustering effect exits in the stock market return?

The structure of this paper is as follows. Section 2 consists of a literature review. Section 3 methodology. Section 4 results and financings, Section 5 discussion and conclusion.

## 2. Literature Review

After the outbreak of the COVID-19 so far research paper has been published about the impact of COVID-19 on the stock market and the world economy before this pandemic there is literature available on the Asian flu 1957 and SARS 2003 which shows that the negative impact on the stock market all over the world.

Barro, R., Ursua, J., & Weng, J. (2020) compared the Spanish Flu losses with COVID-19 and predict the consequences of COVID-19 on the world economy, further they that the COVID-19 has caused to market crashes, volatility, decline in the interest rate and decrease in the economic activities. The research used forecasting model to identify future dividend and its growth, results reveal that the significant decrease in the dividend growth 16% and 23% in the USA and Europe respectively and decline in the growth of GDP expected to 3.6% and 5% in the USA and Europe respectively till 12$^{th}$ of May, 2020, further result reveal that within the two years of period expected dividend growth will -29% and -38% in the USA and Europe respectively (Gormsen, N. J., & Koijen, R. S. 2020).

Cajner, T., Crane, L. D., Decker, R. A., Hamins-Puertolas, A., & Kurz, C. (2020) researched the developments the labor market during COVID-19, the results show that 13 million people lost their jobs in just two weeks from 14 March to 28 March 2020 in the USA, further they compared that 9 million jobs people lost their jobs in the Great Recession in the USA, moreover most effected sector from the COVID-19 is hospitality 30% decline in the employment which is almost 4 million people lost their jobs. Humphries, J. E., Neilson, C., & Ulyssea, G. (2020) surveyed 8 thousand small business owners USA based and they show main three findings in their research, the first one is that the 60% of the business owners already fired one worker from the job as they don't aware with the CARES act by USA government which is also 2$^{nd}$ finding of the research, and the third one is 46% of the business owner that the COVID-19 negative impact will remain for next two years.

Bartik, A. W., Bertrand, M., Cullen, Z. B., Glaeser, E. L., Luca, M., & Stanton, C. T. (2020) has surveyed 5,800 owners of the USA based small business and reveal some important finding that the 43% of the business has been stopped their activities which means they closed their business temporary and the 40% decreased seen in the employment rate, the number of business has been disabled financially and numbers of business are awaited for the Government aids.

Alfaro, L., Chari, A., Greenland, A. N., & Schott, P. K. (2020) researched the change in the market return due to COVID-19 and research shows that 4% to 11 % significantly decline has been seen in the total market value and further they find that the increase in the number of cases causes a decrease in the volatility of the market returns.

Zhang, D., Hu, M., & Ji, Q. (2020) research about the impact of COVID-19 on world financial markets, they argue that the due to COVID-19 record level of risk increase in the market which affected the investors in very limited time. Onali, E. (2020) investigate the COVID-19 effect in term of the number of cases and deaths on US and Europe stock markets and results reveal that there is no impact of COVID-19 on the market returns of US Stock market also he finds that there is a negative relationship between COVID-19 cases and market returns of the Italy and France stock exchange. Nozawa, Y., & Qiu, Y. (2020) investigate the market reaction of the corporate bonds during the COVID-19 and they find that the Central bank promised to support cut down the default risk for loan borrowers and further, the result shows mixed evidence about the market reactions caused by the market segments and liquidity channels.

Ortmann, R., Pelster, M., & Wengerek, S. T. (2020) investigate the impact of COVID-19 outbreak on the retail investors, the findings show that the significant increase has been seen in the stock trading while increase in the cases specifically older age and male investors, 13.9 % trading in stock increased in the week which affects the stock index and 9.99% decline recorded in Dow 30 on $12^{th}$ March 2020.

Liu, H., Manzoor, A., Wang, C., Zhang, L., & Manzoor, Z. (2020) investigate the impact of COVID-19 on world 21 major stock indices in short term, the results show that the word major stock markets have directly affected due to COVID-19 and significantly decreased after the COVID-19 outbreak has been recorded, moreover their results show that Asian countries are more affected than the other regions. Further, regression analyses reveal that there is a negative relationship between the increase in the number of cases and stock indices return.

Çıtak, F., Bagci, B., Şahin, E. E., Hoş, S., & Sakinc, İ. (2020) investigate the effect of COVID-19 on the stock market they found that there is an existence of significant and negative impact of COVID-19 on the Stock market of all the countries. Heyden, K. J., & Heyden, T. (2020) investigate that what impact on the stock market of USA and European after the report of first COVID-19 case in the country, the result shows that there is a negative relationship between the report of the first case and the stock market, further they identify that the fiscal policy also negatively impacted on the stock market returns, improvement in the stock market has recorded after the announcement of monetary policy.

## 3. Research Methodology

Research Method states to technique is being used to perform research related to business; it offers a technique to the examined outcome for particular Challenge in Research Study intended for the whole study is conducting, it shows the path, road-map, combination, and sense for creating dependable results and create outcome beneficial for every stake-holders used for that study, a proper method can Create comprehensive outcomes or vice versa, that's why procedure retains the worth of core part in researches.

In this research the secondary data has been collected of 41 major stock markets indices from the 32 countries data has been collected from investing.com for stock indices from 1$^{st}$ July 2019 to 14$^{th}$ May 2020 and www.ourworldindata.org for COVID19 cases on daily basis from the period of reported the first case of COVID19 according to the country till 14$^{th}$ May 2020. The market has been classified into developed and the emerging market further we divided the data according to the subcontinent (Morgan Stanley Classification Index), below tables represents the classification of indices and countries.

Table 1 Detail of Indexes

| S.no | Codes | Last | Country | MSCI | Continent |
|---|---|---|---|---|---|
| 1 | A | Dow 30 | American | Developed | American |
| 2 | B | S&P 500 | American | Developed | American |
| 3 | CI | Nasdaq | American | Developed | American |
| 4 | DI | SmallCap 2000 | American | Developed | American |
| 5 | E | S&P 500 VIX | American | Developed | American |
| 6 | G | DAX | Germany | Developed | Europe |
| 7 | H | FTSE 100 | UK | Developed | Europe |
| 8 | I | CAC 40 | France | Developed | Europe |
| 9 | J | AEX | Netherland | Developed | Europe |
| 10 | K | IBEX 35 | Spain | Developed | Europe |
| 11 | L | FTSE MIB | Italy | Developed | Europe |
| 12 | M | SMI | Switzerland | Developed | Europe |
| 13 | N | PSI 20 | Portugal | Developed | Europe |
| 14 | O | BEL 20 | Belgium | Developed | Europe |
| 15 | P | ATX | Austria | Developed | Europe |
| 16 | Q | OMXS30 | Sweden | Developed | Europe |
| 17 | R | OMXC25 | Denmark | Developed | Europe |
| 18 | S | Nikkei 225 | Japan | Developed | Pacific |
| 19 | T | S&P/ASX 200 | Australia | Developed | Pacific |
| 20 | U | DJ New Zealand | New Zealand | Developed | Pacific |
| 21 | W | STI Index | Singapore | Developed | Pacific |
| 22 | X | TA 35 | Israel | Developed | Europe |
| 1 | Y | Bovespa | Brazil | Emerging | American |
| 2 | Z | S&P/BMV IPC | Mexico | Emerging | American |
| 3 | AA | MOEX | Russia | Emerging | Europe |
| 4 | AB | RTSI | Russia | Emerging | Europe |
| 5 | AC | WIG20 | Poland | Emerging | Europe |
| 6 | AD | Budapest SE | Hungary | Emerging | Europe |
| 7 | AE | BIST 100 | Turkey | Emerging | Europe |
| 8 | AF | Tadawul All Share | Saudi Arab | Emerging | Middle East |
| 9 | AG | Shanghai | China | Emerging | Asia |
| 10 | AH | SZSE Component | China | Emerging | Asia |
| 11 | AI | China A50 | China | Emerging | Asia |
| 12 | AJ | DJ Shanghai | China | Emerging | Asia |
| 13 | AK | Taiwan Weighted | Taiwan | Emerging | Asia |
| 14 | AL | SET | Thailand | Emerging | Asia |
| 15 | AM | IDX Composite | Indonesia | Emerging | Asia |
| 16 | AN | Nifty 50 | India | Emerging | Asia |
| 17 | AO | BSE Sensex | India | Emerging | Asia |
| 18 | AP | PSEi Composite | Philippine | Emerging | Asia |
| 19 | AQ | Karachi 100 | Pakistan | Emerging | Asia |

The table above, exhibiting Indices Name, Alphabetical Codes, Country, Market, and Region Classification, we have categorized market and region according to the indexes of Morgan Stanley Capital International. This study had encountered 41 the best performing indexes around the globe categorized by Investing.com, further alphabetical codes were used in the model construction.

### 3.1 Research Procedure

### 3.1.1 Independent and Dummy Variables:

In this research we used dummy variable as reported Covid-19 cases in each of the countries since the very first case by assigning as 1 to the Covid-19 reported cases in the EGARCH Equation, however not only we use a dummy variable to measure clustering effects in the model but also we employed cumulative Covid-19 cases in place of dummy variable but this sort of model we not giving logical results so we dropped these model and decided to use dummy variable in the research. In the regression model, we used % change of Covid-19 cases daily in each of the countries then regressed these % changes of Covid-19 cases to the daily indexes returns.

### 3.1.2 Dependent Variables:

Daily market average returns are used to as dependent variable in the regression and individual EGARCH models, and in the classification wise models, we have averaged out indices daily market return concerning their region and market classification and then create a single index such as Index of Developed Market, Emerging Market, American, European, Asian and Pacific & Gulf markets.

### 3.1.3 Hypothesis

**Null Hypothesis**: There is no significant impact on the returns on each selected indices by the Coronavirus Pandemic.

**Alternative Hypothesis**: There is a significant impact on each return on selected indices by the Coronavirus Pandemic.

**Null Hypothesis**: There is no presence of clustering effects in the returns and volatility of each selected indices, Developed and Emerging Markets, and Continental Markets during the Coronavirus Pandemic.

**Alternative Hypothesis**: There is a presence of clustering effects in the returns and volatility of each selected indices, Developed and Emerging Markets, and Continental Markets during the Coronavirus Pandemic.

### 3.1.4 Symbolically Representation of Hypothesis

*Model # 1 Alternative Hypothesis (Regression Mode- Individual Index)*

$H_A : \beta_{returns} = \beta_2 = \beta + \beta_{\%change\ Covid-19}$ *case + er*

$H_B : \beta_{returns} = \beta_2 = \beta + \beta_{\%change\ Covid-19}$ *case + er*

$H_{CI} : \beta_{returns} = \beta_2 = \beta + \beta_{\%change\ Covid-19}$ *case + er*

$H_{DI} : \beta_{returns} = \beta_2 = \beta + \beta_{\%change\ Covid-19}$ *case + er* ………………. $H_{AQ} : \beta_{returns} = \beta_2 = \beta + \beta_{\%change\ Covid-19}$ *case + er*

(All Regression Models from A to AQ )

*Model # 2 Alternative Hypothesis (EGARCH Model-Individual Index)*

$H_0 : \beta_{returns} = \beta_A = \beta_B = \beta_{CI} = \beta_{DI} = \beta_E = \beta_G = \beta_H = \ldots\ldots\ldots\ldots = \beta_{AO} = \beta_{AP} = \beta_{AQ}$

$H_1 : \beta_{returns} \neq \beta_A = \beta_B = \beta_{CI} = \beta_{DI} = \beta_E = \beta_G = \beta_H = \ldots\ldots\ldots\ldots = \beta_{AO} = \beta_{AP} = \beta_{AQ}$

(ALL EGARCH Models from A to AQ)

*Model # 3 Alternative Hypothesis (EGARCH Model-Developed and Emerging Markets)*

$H_0 : \beta_{\text{Indices Average returns}} = \beta_{\text{Developed Markets}} = \beta_{\text{Emerging Markets}}$

$H_1 : \beta_{\text{Indices Average returns}} \neq \beta_{\text{Developed Markets}} = \beta_{\text{Emerging Markets}}$

(EGARCH Models on Developed and Emerging Market by Using Dummy Variables)

*Model # 4 Alternative Hypothesis (EGARCH Model-Continental Markets)*

$H_0 : \beta_{\text{Indices Average returns}} = \beta_{\text{America Market}} = \beta_{\text{Asia Market}} = \beta_{\text{European Market}} = \beta_{\text{Pacific and Gulf Market}}$

$H_1 : \beta_{\text{Indices Average returns}} \neq \beta_{\text{America Market}} = \beta_{\text{Asia Market}} = \beta_{\text{European Market}} = \beta_{\text{Pacific and Gulf Market}}$

(EGARCH Models on Continental Market by Using Dummy Variables)

### 3.1.5 Plan of Analysis / Statistical Tools

In the first step, we examined shifts in the global indices by using descriptive statistics from 1$^{st}$ July 2019 to 14$^{th}$ May 2020 to compare pre and post-pandemic situation, shifts are classified as Median, Standard Deviation, and Relatives Ranking of each index, Median and Standard Deviation is calculated based on daily return and compressed to the monthly returns and then this monthly median and the standard deviation is used to assign relative rankings.

In the second segment, we quantified the Correlation matrix before and after the pandemic to see the international indexes joint movement to each other, is this segment we developed two matrixes and its analysis.

The third phase represents the impact of COVID-19 on indices return in the continent, for that we individually run regression model by using linear regression model using the daily basis data starting from the first case reported of COVID-19 in the country to 14$^{th}$ of May, 2020 and see constructed a comprehensive table which illustrates how each index is effect by the % change in Covid-19 cases by its coefficient and p-values.

Not only we set the regression for each index, but also we see the clustering effect in each index in phase four by using EGARCH model to use of daily basis data from 1st July 2019 to 14th May 2020 and see the clustering effect by dividing them into two broad categories as Developed and Emerging Market as per the guideline of MCSI.

Now in the fifth phase of the research, we constructed a single index by averaging daily returns of respective indices in two broad categories as Developed and Emerging Market, this time we tested the joint clustering effect on Developed and Emerging Markets by assigning it a single index through Averaging.

In the last segment phase 6 we use the same methodology as in step five, but this time we jointly tested the clustering effects in all of the 4 sampled continents by using the EGARCH model.

We have employed Descriptive Statistics, Ordinary Least Square (OLS), Correlation Matrixes, ARCH, GARCH, TARCH, EGARCH and PARCH model, therefore we did not find TARCH and PARCH suitable for this study so we dropped these model, then from GARCH and EGRACH we selected EGARCH model according to the Information Criterion Tests (AIC, HIC, and HQC) and in the last of every EGARCH model, we use ARCH-LM test to diagnose the ARCH type of effect in the model.

OLS, GARCH, TGARCH, and EGARCH models were estimated and the model with the lowest values of AIC, SC, and HQC criterions was selected.

The following EGARCH model was assessed for the study.

### 3.1.6 Exponential Generalized Autoregressive Conditional Heteroskedasticity (EGARCH)

$$\text{Log}(\sigma^2_t) = w + \beta_{log} + (\sigma^2_{t-1}) + \alpha \left\lfloor \frac{E_{t-1}}{\sigma_{t-1}} \right\rfloor + \gamma \frac{E_{t-1}}{\sigma_{t-1}}$$

The equation above denotes the constant of the variance equation; ∐ is the βlog + (σ2t-1) GARCH term which evaluates the size of the group effect in the restricted volatility of the selected indices returns. α ⌊(Et-1)/(σt-1)⌋ is the ARCH term which measures the closeness and scope of ARCH influence in the measured conditional fluctuation. Is the asymmetric γ (Et-1)/(σt-1) expression which evaluates the vastness of asymmetric effect. Asymmetric term measures the size of the uneven effect in the restricted variations of the selected indices return. Adverse innovation, normally principals for the greatest part stimulates higher next period volatility distinguished with positive development. This component is known as Asymmetric impact (Ding et al., 1993).

# 4. Results And Findings

Table 2 Descriptive Statistic for Panel 1 (America)

| Indexes | Descriptive | 31-Jul-19 | 30-Aug-19 | 30-Sep-19 | 31-Oct-19 | 29-Nov-19 | 31-Dec-19 | 31-Jan-20 | 28-Feb-20 | 31-Mar-20 | 30-Apr-20 | 14-May-20 |
|---|---|---|---|---|---|---|---|---|---|---|---|---|
| Dow 30 | Median of Returns | 0.04% | 0.18% | 0.14% | 0.09% | 0.11% | 0.11% | 0.11% | -0.43% | -1.41% | 0.17% | -0.17% |
| | Stdv of Returns | 0.491% | 1.381% | 0.578% | 0.798% | 0.405% | 0.529% | 0.788% | 1.624% | 6.327% | 2.661% | 1.577% |
| | Ranks on Returns | 17 | 6 | 17 | 25 | 18 | 21 | 10 | 40 | 39 | 37 | 31 |
| | Ranks on Stdv | 41 | 10 | 32 | 27 | 40 | 38 | 30 | 18 | 4 | 8 | 19 |
| | No. of Cases | - | - | - | - | - | - | 6 | 60 | 164,620 | 1,039,909 | 1,390,746 |
| S&P 500 | Median of Returns | 0.18% | 0.07% | 0.02% | 0.28% | 0.12% | 0.09% | 0.11% | -0.16% | -1.66% | 0.58% | 0.22% |
| | Stdv of Returns | 0.53% | 1.43% | 0.56% | 0.82% | 0.36% | 0.48% | 0.75% | 1.56% | 5.88% | 2.60% | 1.55% |
| | Ranks on Returns | 5 | 16 | 26 | 9 | 16 | 24 | 10 | 22 | 42 | 22 | 7 |
| | Ranks on Stdv | 38 | 7 | 35 | 24 | 42 | 40 | 34 | 23 | 7 | 11 | 20 |
| | No. of Cases | - | - | - | - | - | - | 6 | 60 | 164,620 | 1,039,909 | 1,390,746 |
| Nasdaq | Median of Returns | 0.20% | -0.11% | -0.09% | 0.33% | 0.17% | 0.20% | 0.14% | 0.11% | -0.70% | 0.77% | 0.85% |
| | Stdv of Returns | 0.69% | 1.63% | 0.85% | 0.93% | 0.48% | 0.54% | 0.87% | 1.74% | 5.73% | 2.62% | 1.69% |
| | Ranks on Returns | 3 | 31 | 40 | 8 | 9 | 12 | 8 | 9 | 30 | 11 | 1 |
| | Ranks on Stdv | 23 | 3 | 12 | 16 | 35 | 37 | 24 | 13 | 8 | 10 | 16 |
| | No. of Cases | - | - | - | - | - | - | 6 | 60 | 164,620 | 1,039,909 | 1,390,746 |
| SmallCap 2000 | Median of Returns | 0.05% | -0.19% | -0.07% | 0.13% | 0.14% | 0.13% | -0.09% | -0.24% | -1.23% | 1.26% | -0.18% |
| | Stdv of Returns | 0.74% | 1.59% | 1.03% | 0.89% | 0.70% | 0.50% | 0.77% | 1.55% | 6.79% | 3.73% | 2.41% |
| | Ranks on Returns | 15 | 35 | 37 | 23 | 14 | 17 | 33 | 27 | 36 | 3 | 32 |
| | Ranks on Stdv | 19 | 4 | 8 | 18 | 16 | 39 | 31 | 24 | 3 | 2 | 2 |
| | No. of Cases | - | - | - | - | - | - | 6 | 60 | 164,620 | 1,039,909 | 1,390,746 |
| S&P 500 VIX | Median of Returns | 0.08% | -2.11% | -2.12% | -1.75% | 0.63% | -0.16% | 0.16% | 0.93% | -2.24% | -3.33% | -2.38% |
| | Stdv of Returns | 6.51% | 14.41% | 6.33% | 7.43% | 4.29% | 7.45% | 9.13% | 15.18% | 17.94% | 7.28% | 9.53% |
| | Ranks on Returns | 13 | 43 | 43 | 43 | 2 | 43 | 3 | 1 | 43 | 43 | 43 |
| | Ranks on Stdv | 1 | 1 | 1 | 1 | 1 | 1 | 1 | 1 | 1 | 1 | 1 |
| | No. of Cases | - | - | - | - | - | - | 6 | 60 | 164,620 | 1,039,909 | 1,390,746 |
| Bovespa | Median of Returns | 0.12% | 0.43% | 0.21% | 0.35% | -0.16% | 0.33% | -0.26% | -0.50% | -1.44% | 1.37% | -0.32% |
| | Stdv of Returns | 0.79% | 1.54% | 0.66% | 1.12% | 0.90% | 0.62% | 1.35% | 1.99% | 7.69% | 2.95% | 1.52% |
| | Ranks on Returns | 10 | 1 | 13 | 2 | 40 | 3 | 39 | 41 | 40 | 1 | 36 |
| | Ranks on Stdv | 15 | 5 | 25 | 3 | 8 | 27 | 4 | 8 | 2 | 4 | 21 |
| | No. of Cases | - | - | - | - | - | - | - | 1 | 4,579 | 78,162 | 188,974 |
| S&P/BMV IPC | Median of Returns | -0.16% | 0.27% | -0.09% | -0.08% | 0.01% | -0.09% | -0.07% | -0.28% | -1.33% | 0.13% | -0.13% |
| | Stdv of Returns | 0.84% | 1.20% | 0.67% | 0.85% | 0.68% | 0.94% | 1.07% | 1.13% | 3.16% | 1.74% | 1.40% |
| | Ranks on Returns | 42 | 2 | 41 | 37 | 28 | 42 | 30 | 31 | 38 | 38 | 29 |
| | Ranks on Stdv | 12 | 16 | 24 | 21 | 19 | 4 | 13 | 38 | 34 | 30 | 24 |
| | No. of Cases | - | - | - | - | - | - | - | - | 1,094 | 17,799 | 40,186 |

The purpose of keeping above mentioned statistic was to analyze the median shift in the returns of the each index, therefore not only we reported the shift in the median as well as encountered the risk associated with the return in term of standard deviation it has been measure, shift of median showcased that since Novel Covid-19 has not been declared as pandemic, the indexes above exhibiting the strong stability with steady risk associated, on 11 March Novel Covid-19 declared as pandemic by the World Health Organization (WHO), then stock market aggressively shown the abnormal change and huge shift in the average returns, we have target the indexes classified into American region, whereas the virus largely hit the world biggest economy so perceive consequences in term of stock markets could be seen into American stock market, therefore the shift of median from month to month trailing 11 months encountered in this research exhibited massive change each American index, Covid-19 declared pandemic outbreak in the month of March and the massive change have been detected into each American indexes, the best ranked index become eventually the worsen such as Dow 30, S&P 500, Nasdaq, SmallCap 2000 and S&P 500 VIX ranked 39, 42, 30, 36 and 43 in Mar20 (Jul-19, 17, 5, 3, 15 and 13) on basis of monthly median returns respectively, further we had ranked entire indices classified by the investing.com, hence S&P/TSX (Canada) and Hang Seng (Hong Kong) indexes were removed due to insufficient data of Covid-19, moving forward in term of standard deviation the highest ranked indexed best and vice versa, meaning indices with bigger rank number are consider best because risk associated with them are on very low level therefore comparing to this phenomena again American indexes reported the worsen ranked in term of risk, such as one to five indexes shown the shift in risk 8, 11, 10, 2 and 1 in Mar20 (Jul-19 41, 38, 23, 19, 1) respectively. The information also claims, when Covid-19 cases shown significant increment in American ultimately put the impact on the American stock market and stood the stock market on the verge of collapse. Therefore the American stock market was classified as the highest vulnerable stock market around the globe. Not only there is a significant shift into the rank but also the aggressive shift has been observed into the American stock market within 5 months only especially in the month of March-2020 after declaration as a pandemic.

Table 3 Descriptive Statistic for Panel 2 (Asia)

| Indexes | Descriptive | 31-Jul-19 | 30-Aug-19 | 30-Sep-19 | 31-Oct-19 | 29-Nov-19 | 31-Dec-19 | 31-Jan-20 | 28-Feb-20 | 31-Mar-20 | 30-Apr-20 | 14-May-20 |
|---|---|---|---|---|---|---|---|---|---|---|---|---|
| Shanghai | Median of Returns | 0.03% | -0.11% | 0.23% | -0.35% | -0.07% | 0.24% | -0.52% | 0.31% | -0.47% | 0.25% | 0.19% |
| | Stdv of Returns | 0.93% | 0.97% | 0.75% | 0.64% | 0.69% | 0.68% | 0.94% | 2.17% | 1.85% | 0.87% | 0.51% |
| | Ranks on Returns | 21 | 29 | 10 | 40 | 33 | 8 | 42 | 4 | 22 | 35 | 9 |
| | Ranks on Stdv | 9 | 33 | 16 | 35 | 18 | 24 | 19 | 5 | 43 | 43 | 43 |
| | No. of Cases | - | - | - | - | - | 27 | 9,714 | 78,927 | 82,241 | 83,944 | 84,024 |
| SZSE Component | Median of Returns | 0.15% | -0.11% | 0.30% | -0.31% | -0.06% | 0.40% | -0.17% | 0.58% | -0.47% | 0.28% | 0.48% |
| | Stdv of Returns | 1.26% | 1.20% | 1.07% | 0.83% | 0.98% | 0.88% | 1.31% | 2.80% | 2.42% | 1.21% | 0.74% |
| | Ranks on Returns | 6 | 32 | 5 | 39 | 32 | 2 | 37 | 3 | 21 | 34 | 4 |
| | Ranks on Stdv | 3 | 17 | 6 | 23 | 4 | 8 | 5 | 2 | 40 | 39 | 37 |
| | No. of Cases | - | - | - | - | - | 27 | 9,714 | 78,927 | 82,241 | 83,944 | 84,024 |
| China A50 | Median of Returns | -0.09% | -0.01% | -0.01% | 0.07% | -0.12% | 0.15% | -0.55% | -0.04% | -0.49% | 0.06% | 0.02% |
| | Stdv of Returns | 1.00% | 1.09% | 0.74% | 0.62% | 0.86% | 0.70% | 0.99% | 2.08% | 2.11% | 0.87% | 0.55% |
| | Ranks on Returns | 36 | 23 | 32 | 27 | 38 | 14 | 43 | 16 | 23 | 40 | 20 |
| | Ranks on Stdv | 4 | 24 | 17 | 37 | 9 | 22 | 16 | 6 | 41 | 42 | 41 |
| | No. of Cases | - | - | - | - | - | 27 | 9,714 | 78,927 | 82,241 | 83,944 | 84,024 |
| DJ Shanghai | Median of Returns | -0.01% | -0.11% | 0.13% | 0.00% | -0.11% | 0.24% | 0.00% | 0.25% | -0.40% | 0.30% | 0.00% |
| | Stdv of Returns | 0.96% | 1.03% | 0.77% | 0.63% | 0.74% | 0.70% | 0.92% | 2.22% | 1.97% | 0.89% | 0.52% |
| | Ranks on Returns | 26 | 29 | 20 | 32 | 37 | 8 | 23 | 5 | 19 | 33 | 21 |
| | Ranks on Stdv | 6 | 28 | 14 | 36 | 14 | 21 | 21 | 4 | 42 | 41 | 42 |
| | No. of Cases | - | - | - | - | - | 27 | 9,714 | 78,927 | 82,241 | 83,944 | 84,024 |
| Taiwan Weighted | Median of Returns | -0.07% | 0.04% | 0.07% | 0.17% | 0.08% | 0.13% | 0.00% | -0.29% | -0.86% | 0.46% | 0.52% |
| | Stdv of Returns | 0.57% | 0.86% | 0.45% | 0.60% | 0.67% | 0.57% | 1.40% | 0.95% | 3.02% | 1.27% | 1.21% |
| | Ranks on Returns | 31 | 18 | 23 | 19 | 21 | 17 | 25 | 32 | 32 | 26 | 2 |
| | Ranks on Stdv | 32 | 38 | 39 | 40 | 20 | 33 | 2 | 41 | 38 | 38 | 26 |
| | No. of Cases | - | - | - | - | - | - | 9 | 34 | 306 | 429 | 440 |
| SET | Median of Returns | -0.02% | -0.28% | 0.04% | -0.08% | -0.18% | -0.07% | -0.04% | -0.42% | 0.49% | 0.69% | 0.38% |
| | Stdv of Returns | 0.48% | 0.91% | 0.54% | 0.60% | 0.69% | 0.60% | 1.02% | 1.87% | 4.51% | 2.06% | 1.16% |
| | Ranks on Returns | 28 | 37 | 24 | 37 | 41 | 39 | 29 | 39 | 5 | 17 | 5 |
| | Ranks on Stdv | 42 | 36 | 37 | 38 | 17 | 28 | 14 | 10 | 21 | 25 | 28 |
| | No. of Cases | - | - | - | - | - | - | 14 | 40 | 1,651 | 2,954 | 3,017 |
| IDX Composite | Median of Returns | 0.02% | 0.02% | -0.05% | 0.19% | -0.31% | 0.21% | -0.08% | -0.34% | -1.49% | 0.38% | -0.36% |
| | Stdv of Returns | 0.54% | 0.86% | 0.70% | 0.72% | 0.62% | 0.58% | 0.73% | 0.94% | 4.01% | 2.08% | 0.98% |
| | Ranks on Returns | 23 | 20 | 36 | 14 | 43 | 11 | 31 | 36 | 41 | 32 | 37 |
| | Ranks on Stdv | 36 | 39 | 21 | 34 | 26 | 31 | 35 | 42 | 25 | 24 | 35 |
| | No. of Cases | - | - | - | - | - | - | - | - | 1,414 | 9,771 | 15,438 |
| Nifty 50 | Median of Returns | -0.15% | 0.17% | 0.00% | 0.14% | 0.08% | 0.08% | -0.03% | -0.27% | -0.54% | 0.76% | -0.30% |
| | Stdv of Returns | 0.74% | 1.02% | 1.63% | 0.75% | 0.52% | 0.56% | 0.77% | 1.26% | 4.84% | 2.81% | 2.21% |
| | Ranks on Returns | 41 | 8 | 29 | 21 | 21 | 26 | 28 | 29 | 27 | 12 | 33 |
| | Ranks on Stdv | 20 | 30 | 2 | 33 | 32 | 35 | 32 | 33 | 13 | 6 | 5 |
| | No. of Cases | - | - | - | - | - | - | 1 | 3 | 1,251 | 33,050 | 78,003 |
| BSE Sensex | Median of Returns | -0.09% | 0.20% | -0.04% | 0.19% | 0.07% | 0.02% | 0.03% | -0.26% | -0.50% | 0.73% | -0.43% |
| | Stdv of Returns | 0.70% | 1.01% | 1.63% | 0.82% | 0.52% | 0.57% | 0.75% | 1.23% | 4.96% | 2.90% | 2.28% |
| | Ranks on Returns | 36 | 5 | 35 | 14 | 26 | 32 | 19 | 28 | 24 | 14 | 39 |
| | Ranks on Stdv | 22 | 31 | 3 | 25 | 33 | 34 | 33 | 34 | 11 | 5 | 4 |
| | No. of Cases | - | - | - | - | - | - | 1 | 3 | 1,251 | 33,050 | 78,003 |
| PSEi Composite | Median of Returns | 0.03% | 0.11% | -0.03% | 0.19% | -0.14% | 0.04% | -0.15% | -0.05% | 0.05% | 0.46% | -0.02% |
| | Stdv of Returns | 0.93% | 1.18% | 0.57% | 0.85% | 0.95% | 0.97% | 1.07% | 1.52% | 4.84% | 2.66% | 1.07% |
| | Ranks on Returns | 22 | 12 | 34 | 14 | 39 | 29 | 36 | 18 | 12 | 26 | 25 |
| | Ranks on Stdv | 8 | 18 | 34 | 22 | 5 | 3 | 12 | 26 | 14 | 9 | 31 |
| | No. of Cases | - | - | - | - | - | - | 1 | 3 | 2,084 | 8,212 | 11,618 |
| Karachi 100 | Median of Returns | 0.04% | -1.44% | -0.01% | 0.35% | 0.69% | 0.22% | -0.13% | -0.52% | -0.53% | 0.74% | 0.14% |
| | Stdv of Returns | 0.97% | 1.72% | 1.05% | 0.95% | 1.27% | 1.16% | 1.21% | 1.28% | 3.31% | 2.27% | 0.69% |
| | Ranks on Returns | 18 | 42 | 31 | 2 | 1 | 10 | 35 | 42 | 26 | 13 | 13 |
| | Ranks on Stdv | 5 | 2 | 7 | 12 | 2 | 2 | 9 | 32 | 33 | 17 | 40 |
| | No. of Cases | - | - | - | - | - | - | - | 2 | 1,625 | 15,759 | 35,788 |

Many articles classified Asian stock market as the least vulnerable market, hence this is being exhibited by the above-mentioned table, an aggressive is seen only in Chinese stock market and rest market reported steady risk and stable returns, going towards Chinese stock market which showcases the massive median shift in the month of February-2020 in each Chinese indexes, it is very interesting before pandemic declaration the Chinese market more vulnerable and later proclaimed the significant stability in term of Monthly Average Returns, Standard Deviation, Median and Standard Deviation ranks. The number indicates that when there is a massive change in the number of Covid-19 reported cases in January and February then the market lost its stability. Coming towards rest of the stock market into Asia, Indian stock market shows healthy improvement in term of average monthly returns and stable rank shift, not only Indian Stock market but also Pakistani stock market shows stability in the global pandemic because these are the countries which adopted smart lockdown policy and intended to run stock market as usual.

Table 4 Descriptive Statistic for Panel 3 (Europe)

| Indexes | Descriptive | 31-Jul-19 | 30-Aug-19 | 30-Sep-19 | 31-Oct-19 | 29-Nov-19 | 31-Dec-19 | 31-Jan-20 | 28-Feb-20 | 31-Mar-20 | 30-Apr-20 | 14-May-20 |
|---|---|---|---|---|---|---|---|---|---|---|---|---|
| DAX | Median of Returns | 0.01% | 0.11% | 0.32% | 0.34% | 0.10% | -0.07% | -0.02% | -0.03% | -0.31% | 0.95% | -0.39% |
|  | Stdv of Returns | 0.81% | 1.26% | 0.55% | 1.04% | 0.51% | 0.78% | 1.00% | 1.48% | 4.66% | 2.57% | 1.97% |
|  | Ranks on Returns | 24 | 11 | 4 | 5 | 19 | 39 | 26 | 15 | 16 | 8 | 38 |
|  | Ranks on Stdv | 13 | 13 | 36 | 4 | 34 | 15 | 15 | 29 | 19 | 12 | 9 |
|  | No. of Cases | - | - | - | - | - | - | 5 | 47 | 61,913 | 159,119 | 172,239 |
| FTSE 100 | Median of Returns | -0.05% | -0.07% | -0.02% | 0.09% | 0.13% | 0.11% | 0.01% | -0.27% | 0.86% | 0.55% | -0.05% |
|  | Stdv of Returns | 0.66% | 1.03% | 0.52% | 0.91% | 0.61% | 0.85% | 0.83% | 1.48% | 4.33% | 2.26% | 1.51% |
|  | Ranks on Returns | 30 | 27 | 33 | 25 | 15 | 21 | 21 | 29 | 1 | 23 | 27 |
|  | Ranks on Stdv | 26 | 27 | 38 | 17 | 27 | 11 | 28 | 30 | 22 | 19 | 22 |
|  | No. of Cases | - | - | - | - | - | - | 2 | 16 | 22,141 | 165,221 | 229,705 |
| CAC 40 | Median of Returns | -0.03% | 0.16% | 0.23% | 0.17% | 0.08% | 0.13% | -0.02% | -0.19% | 0.42% | 0.65% | -0.75% |
|  | Stdv of Returns | 0.57% | 1.42% | 0.62% | 1.02% | 0.38% | 0.78% | 0.89% | 1.59% | 4.75% | 2.36% | 2.02% |
|  | Ranks on Returns | 29 | 9 | 8 | 19 | 21 | 17 | 26 | 23 | 6 | 19 | 41 |
|  | Ranks on Stdv | 35 | 8 | 29 | 6 | 41 | 14 | 23 | 19 | 16 | 14 | 7 |
|  | No. of Cases | - | - | - | - | - | - | 6 | 38 | 44,550 | 128,442 | 140,734 |
| AEX | Median of Returns | 0.14% | 0.13% | 0.20% | -0.02% | 0.15% | -0.01% | 0.07% | 0.01% | 0.35% | 1.23% | -0.02% |
|  | Stdv of Returns | 0.53% | 1.23% | 0.44% | 0.94% | 0.47% | 0.79% | 0.92% | 1.74% | 4.23% | 2.04% | 1.90% |
|  | Ranks on Returns | 8 | 10 | 15 | 34 | 13 | 35 | 14 | 13 | 7 | 4 | 24 |
|  | Ranks on Stdv | 39 | 15 | 41 | 14 | 36 | 13 | 22 | 14 | 23 | 26 | 11 |
|  | No. of Cases | - | - | - | - | - | - | - | 1 | 11,750 | 38,802 | 43,211 |
| IBEX 35 | Median of Returns | -0.12% | 0.21% | 0.29% | 0.03% | -0.05% | 0.02% | -0.17% | 0.05% | -0.07% | 0.40% | -0.30% |
|  | Stdv of Returns | 0.807% | 1.023% | 0.629% | 0.973% | 0.573% | 0.866% | 0.822% | 1.700% | 4.917% | 2.096% | 1.656% |
|  | Ranks on Returns | 39 | 4 | 6 | 30 | 31 | 32 | 37 | 11 | 14 | 30 | 34 |
|  | Ranks on Stdv | 14 | 29 | 28 | 9 | 31 | 9 | 29 | 17 | 12 | 23 | 18 |
|  | No. of Cases | - | - | - | - | - | - | - | 35 | 104,267 | 215,183 | 228,691 |
| FTSE MIB | Median of Returns | 0.06% | -0.22% | 0.15% | 0.21% | 0.07% | 0.14% | 0.02% | 0.12% | 0.32% | 0.95% | 0.09% |
|  | Stdv of Returns | 0.95% | 1.48% | 0.69% | 0.95% | 0.66% | 0.86% | 1.18% | 1.90% | 5.42% | 2.33% | 1.81% |
|  | Ranks on Returns | 14 | 36 | 16 | 13 | 25 | 16 | 20 | 8 | 8 | 9 | 16 |
|  | Ranks on Stdv | 7 | 6 | 23 | 13 | 22 | 10 | 10 | 9 | 10 | 15 | 13 |
|  | No. of Cases | - | - | - | - | - | - | 3 | 650 | 101,739 | 203,591 | 222,104 |
| SMI | Median of Returns | 0.09% | -0.12% | 0.01% | 0.25% | 0.19% | 0.17% | 0.10% | 0.03% | 0.58% | 0.40% | 0.35% |
|  | Stdv of Returns | 0.62% | 0.97% | 0.58% | 0.77% | 0.45% | 0.68% | 0.71% | 1.57% | 3.78% | 1.42% | 1.20% |
|  | Ranks on Returns | 12 | 33 | 28 | 12 | 8 | 13 | 13 | 12 | 3 | 31 | 6 |
|  | Ranks on Stdv | 28 | 34 | 33 | 30 | 37 | 23 | 36 | 22 | 28 | 34 | 27 |
| PSI 20 | No. of Cases | - | - | - | - | - | - | - | 8 | 15,412 | 29,324 | 30,330 |
| | Median of Returns | -0.28% | 0.01% | 0.22% | 0.19% | -0.08% | 0.01% | 0.04% | -0.15% | 0.18% | 0.59% | -0.77% |
| | Stdv of Returns | 0.68% | 1.10% | 0.75% | 0.60% | 0.61% | 0.62% | 0.68% | 1.49% | 4.01% | 1.49% | 1.37% |
| | Ranks on Returns | 43 | 21 | 11 | 14 | 34 | 34 | 18 | 21 | 10 | 21 | 42 |
| | Ranks on Stdv | 25 | 23 | 15 | 39 | 28 | 26 | 38 | 27 | 26 | 33 | 25 |
| BEL 20 | No. of Cases | - | - | - | - | - | - | - | - | 6,408 | 24,692 | 28,132 |
| | Median of Returns | 0.14% | -0.04% | 0.38% | 0.10% | 0.17% | 0.10% | -0.08% | -0.04% | 0.21% | 1.21% | 0.12% |
| | Stdv of Returns | 0.76% | 1.25% | 0.60% | 1.00% | 0.45% | 0.58% | 0.87% | 2.02% | 4.72% | 2.47% | 2.15% |
| | Ranks on Returns | 7 | 25 | 3 | 24 | 11 | 23 | 31 | 16 | 9 | 5 | 14 |
| | Ranks on Stdv | 16 | 14 | 31 | 8 | 38 | 32 | 25 | 7 | 17 | 13 | 6 |
| ATX | No. of Cases | - | - | - | - | - | - | - | 1 | 11,899 | 47,859 | 53,981 |
| | Median of Returns | -0.15% | -0.32% | 0.21% | 0.27% | 0.08% | 0.11% | -0.12% | -0.35% | -0.61% | 1.00% | -0.74% |
| | Stdv of Returns | 0.73% | 1.00% | 0.70% | 0.93% | 0.78% | 0.59% | 0.71% | 1.55% | 5.55% | 2.75% | 1.94% |
| | Ranks on Returns | 40 | 40 | 12 | 10 | 21 | 20 | 34 | 37 | 28 | 7 | 40 |
| | Ranks on Stdv | 21 | 32 | 22 | 15 | 13 | 30 | 37 | 25 | 9 | 7 | 10 |
| OMXS30 | No. of Cases | - | - | - | - | - | - | - | 5 | 9,618 | 15,364 | 15,964 |
| | Median of Returns | -0.09% | 0.00% | 0.45% | 0.35% | -0.10% | 0.06% | 0.07% | -0.02% | 0.01% | 0.43% | 0.02% |
| | Stdv of Returns | 0.91% | 1.29% | 0.71% | 0.96% | 0.65% | 0.75% | 0.95% | 1.70% | 3.87% | 2.30% | 2.29% |
| | Ranks on Returns | 36 | 22 | 2 | 2 | 36 | 28 | 14 | 14 | 13 | 28 | 19 |
| | Ranks on Stdv | 10 | 12 | 20 | 11 | 23 | 18 | 18 | 16 | 27 | 16 | 3 |
| OMXC25 | No. of Cases | - | - | - | - | - | - | - | 7 | 4,028 | 20,302 | 27,909 |
| | Median of Returns | 0.05% | 0.03% | 0.03% | 0.14% | 0.24% | 0.08% | 0.22% | -0.19% | 0.69% | 0.72% | 0.17% |
| | Stdv of Returns | 0.75% | 1.15% | 0.72% | 0.96% | 0.85% | 0.64% | 0.92% | 1.58% | 3.03% | 1.13% | 1.12% |
| | Ranks on Returns | 15 | 19 | 25 | 21 | 4 | 27 | 2 | 23 | 2 | 15 | 10 |
| | Ranks on Stdv | 17 | 20 | 19 | 10 | 10 | 25 | 20 | 21 | 37 | 40 | 30 |
| TA 35 | No. of Cases | - | - | - | - | - | - | - | 1 | 2,577 | 9,008 | 10,667 |
| | Median of Returns | -0.08% | 0.26% | 0.28% | -0.02% | -0.01% | -0.07% | 0.11% | 0.61% | -1.10% | 0.47% | -0.08% |
| | Stdv of Returns | 0.61% | 1.12% | 0.45% | 0.53% | 0.42% | 0.37% | 0.55% | 1.49% | 3.58% | 1.84% | 1.85% |
| | Ranks on Returns | 35 | 3 | 7 | 34 | 29 | 38 | 10 | 2 | 34 | 25 | 28 |
| | Ranks on Stdv | 29 | 22 | 40 | 42 | 39 | 41 | 41 | 28 | 30 | 29 | 12 |
| MOEX | No. of Cases | - | - | - | - | - | - | - | 3 | 4,473 | 15,834 | 16,548 |
| | Median of Returns | -0.01% | -0.03% | -0.19% | 0.19% | 0.19% | 0.25% | 0.16% | -0.29% | -0.32% | 0.64% | -0.02% |
| | Stdv of Returns | 0.66% | 0.90% | 0.64% | 0.80% | 0.64% | 0.56% | 0.85% | 1.59% | 4.03% | 1.88% | 0.74% |
| | Ranks on Returns | 27 | 24 | 42 | 14 | 7 | 7 | 3 | 32 | 18 | 20 | 26 |
| | Ranks on Stdv | 27 | 37 | 26 | 28 | 25 | 36 | 26 | 20 | 24 | 28 | 38 |
| RTSI | No. of Cases | | | | | | | | | | | |
| | Median of Returns | -0.08% | 0.09% | 0.01% | 0.45% | -0.02% | 0.28% | 0.23% | -0.56% | -0.66% | 1.35% | 0.15% |

| | | | | | | | | | | | |
|---|---|---|---|---|---|---|---|---|---|---|---|
| | Returns | | | | | | | | | | |
| | Stdv of Returns | 0.68% | 1.38% | 0.87% | 0.79% | 0.82% | 0.76% | 1.31% | 2.40% | 5.90% | 3.21% | 1.74% |
| | Ranks on Returns | 34 | 15 | 27 | 1 | 30 | 5 | 1 | 43 | 29 | 2 | 12 |
| | Ranks on Stdv | 24 | 11 | 10 | 29 | 11 | 16 | 6 | 3 | 6 | 3 | 15 |
| | No. of Cases | - | - | - | - | - | - | - | 2 | 1,836 | 99,399 | 242,271 |
| WIG20 | Median of Returns | -0.07% | -0.31% | 0.13% | 0.05% | -0.31% | 0.09% | 0.01% | -0.36% | -0.86% | 0.89% | -0.31% |
| | Stdv of Returns | 0.57% | 1.42% | 0.98% | 1.03% | 0.91% | 0.90% | 1.23% | 1.79% | 4.71% | 2.22% | 1.67% |
| | Ranks on Returns | 31 | 39 | 18 | 29 | 42 | 24 | 21 | 38 | 31 | 10 | 35 |
| | Ranks on Stdv | 34 | 9 | 9 | 5 | 7 | 7 | 8 | 12 | 18 | 20 | 17 |
| | No. of Cases | - | - | - | - | - | - | - | - | 2,055 | 12,640 | 17,204 |
| Budapest SE | Median of Returns | 0.11% | -0.30% | 0.23% | 0.27% | 0.07% | 0.27% | -0.38% | 0.25% | -0.10% | 0.71% | 0.12% |
| | Stdv of Returns | 0.57% | 0.81% | 0.74% | 1.01% | 0.78% | 0.91% | 0.98% | 1.81% | 4.55% | 2.20% | 1.00% |
| | Ranks on Returns | 11 | 38 | 9 | 10 | 27 | 6 | 41 | 5 | 15 | 16 | 14 |
| | Ranks on Stdv | 33 | 40 | 18 | 7 | 12 | 5 | 17 | 11 | 20 | 21 | 34 |
| | No. of Cases | - | - | - | - | - | - | - | - | 492 | 2,775 | 3,380 |
| BIST 100 | Median of Returns | 0.38% | -0.09% | 0.49% | -0.35% | 0.23% | 0.29% | 0.05% | -0.32% | -0.32% | 0.49% | -0.14% |
| | Stdv of Returns | 1.374% | 1.084% | 1.122% | 1.637% | 0.912% | 0.704% | 1.385% | 1.735% | 3.348% | 1.379% | 1.069% |
| | Ranks on Returns | 1 | 28 | 1 | 40 | 5 | 4 | 17 | 34 | 17 | 24 | 30 |
| | Ranks on Stdv | 2 | 25 | 5 | 2 | 6 | 20 | 3 | 15 | 32 | 35 | 32 |
| | No. of Cases | - | - | - | - | - | - | - | - | 10,827 | 117,589 | 143,114 |

The table above comprised on the European Stock Market, hence the Worlddometer show the daily real time data, and the number indicates the destructions of the virus into Europe, so it is the second largest market after America which is succumbed of the pandemic, on today's date (May-15-2020) Russia, Spain, United Kingdom, Italy, France, Germany and Turkey ranked in the top ten effected countries by Covid-19 according to worlddometere. Median & Standard Deviation shift for these indexes are classified as MOEX -0.32% & 4.03% Mar-20 (Jul-19 -0.01 & 0.66), RTSI -0.66% & 5.90% Mar-20 (Jul-19 -0.08% & 0.68%), IBEX 35 -0.07% & 4.91% Mar-20 (Jul-19 -0.12% & 0.80%), FTSE-100 0.86% & 4.33% Mar-20 (Jul-19 -0.05% & 0.66%), FTSE MIB 0.32% & 5.42% Mar-20 (Jul-19 0.06% & 0.95%), CAC-40 0.42% & 4.75% Mar-20 (Jul-19 -0.03% & 0.57%), DAX 0.31% & 4.66% Mar-20 (Jul-19 0.01% & 0.81%) and BIST-100 -0.32% & 3.34% Mar-20 (Jul-19 0.38% & 1.374%). The information concluded that aggressive increment into the Covid-19 cases reflects into stock market of any country, Italy and Spain are quantified the highest vulnerable indexes into Europe due the worse spread of pandemic and complete lock down for many days.

Table 5 Descriptive Statistic for Panel 4 (Pacific & Gulf)

| Indexes | Descriptive | 31-Jul-19 | 30-Aug-19 | 30-Sep-19 | 31-Oct-19 | 29-Nov-19 | 31-Dec-19 | 31-Jan-20 | 28-Feb-20 | 31-Mar-20 | 30-Apr-20 | 14-May-20 |
|---|---|---|---|---|---|---|---|---|---|---|---|---|
| Nikkei 225 | Median of Returns | 0.03% | 0.06% | 0.20% | 0.34% | 0.17% | -0.08% | 0.07% | -0.23% | -1.13% | -0.04% | 0.06% |
| | Stdv of Returns | 0.90% | 0.96% | 0.63% | 0.76% | 0.65% | 0.76% | 1.14% | 1.47% | 3.56% | 2.18% | 1.46% |
| | Ranks on Returns | 20 | 17 | 14 | 5 | 11 | 41 | 14 | 26 | 35 | 42 | 18 |
| | Ranks on Stdv | 11 | 35 | 27 | 31 | 24 | 17 | 11 | 31 | 31 | 22 | 23 |
| | No. of Cases | - | - | - | - | - | - | 14 | 210 | 1,953 | 14,088 | 16,079 |
| S&P/ASX 200 | Median of Returns | 0.32% | 0.18% | 0.13% | 0.07% | 0.21% | -0.04% | 0.13% | -0.14% | -1.24% | 0.00% | -0.02% |
| | Stdv of Returns | 0.51% | 1.12% | 0.41% | 0.81% | 0.67% | 0.90% | 0.62% | 1.18% | 4.83% | 1.96% | 1.98% |
| | Ranks on Returns | 2 | 6 | 18 | 27 | 6 | 36 | 9 | 20 | 37 | 41 | 23 |
| | Ranks on Stdv | 40 | 21 | 42 | 26 | 21 | 6 | 40 | 37 | 15 | 27 | 8 |
| | No. of Cases | - | - | - | - | - | - | 7 | 23 | 4,557 | 6,746 | 6,975 |
| DJ New Zealand | Median of Returns | 0.18% | -0.05% | -0.07% | -0.03% | 0.30% | -0.04% | 0.16% | -0.07% | -0.50% | 0.19% | 0.50% |
| | Stdv of Returns | 0.598% | 1.042% | 0.858% | 0.877% | 0.605% | 0.595% | 0.537% | 1.023% | 3.116% | 1.573% | 0.696% |
| | Ranks on Returns | 4 | 26 | 37 | 36 | 3 | 36 | 3 | 19 | 25 | 36 | 3 |
| | Ranks on Stdv | 31 | 26 | 11 | 20 | 29 | 29 | 42 | 40 | 36 | 32 | 39 |
| | No. of Cases | - | - | - | - | - | - | - | 1 | 647 | 1,129 | 1,147 |
| STI Index | Median of Returns | -0.07% | -0.36% | -0.01% | 0.34% | -0.09% | 0.04% | 0.00% | -0.19% | -0.95% | 0.09% | 0.00% |
| | Stdv of Returns | 0.60% | 0.69% | 0.60% | 0.59% | 0.60% | 0.37% | 0.64% | 1.18% | 3.63% | 1.70% | 1.05% |
| | Ranks on Returns | 31 | 41 | 30 | 5 | 35 | 29 | 23 | 23 | 33 | 39 | 21 |
| | Ranks on Stdv | 30 | 43 | 30 | 41 | 30 | 42 | 39 | 36 | 29 | 31 | 33 |
| | No. of Cases | - | - | - | - | - | - | 13 | 96 | 844 | 15,641 | 25,346 |
| Tadawul All Share | Median of Returns | 0.13% | 0.09% | 0.07% | -0.42% | 0.17% | 0.44% | 0.15% | -0.32% | 0.15% | 0.69% | 0.16% |
| | Stdv of Returns | 0.54% | 0.76% | 0.84% | 0.89% | 0.72% | 0.72% | 0.84% | 0.89% | 3.13% | 1.34% | 0.92% |
| | Ranks on Returns | 9 | 13 | 22 | 42 | 10 | 1 | 7 | 34 | 11 | 17 | 11 |
| | Ranks on Stdv | 37 | 42 | 13 | 19 | 15 | 19 | 27 | 43 | 35 | 36 | 36 |
| | No. of Cases | - | - | - | - | - | - | - | - | 1,453 | 21,402 | 44,830 |

Panel 4 consists on Pacific and Gulf indexes, gulf region countries are least victim countries, business activities are impacted heavily due to coronavirus pandemic in the pacific region and the shift of above-mentioned factors support the statement, the major shift into the factors is being seen in Nikkei 225 (Japan) and STI Index (Singapore) in March. Nikkei 225 reports median average return -1.13% in Mar-20 (Jul-19 0.03%) and Median rank 35 in Mar-20 (Jul-19 20), however, STI Index exhibited monthly average return at -0.95% in Mar-20 (Jul-19 -0.07%) with ranking as 33 in Mar-20 (Jul-19 31), in term of ranks STI is not hit as much higher as Japan, the most considerable thing in the pandemic, according to survey Covid-19 attacks quickly on people who are above 40 years, hence a noticeable thing that Japan inherent skewed population in term of age group, there are more aged people compare to the teenage or young, further the people in an age of 30 to 35 are in established phase and they are potential investors, therefore it can be perceived that Japan's market was given hit due to withdrawal of potential aged investors.

**Table 6 Correlation matrix before Coronavirus**

We have taken the date before and after the pandemic, hence above mentioned correlation matrix are before Coronavirus cases, there are few indexes which have strongly significant correlation such as American and European indexes, therefore the purpose was to see the relationship of the global indexes to each other, hence in the second correlation matrix, we found very different results.

**Table 7 Correlation matrix after Coronavirus**

The European region is considered the most effective countries by Covid-19, above matrix, was constructed after the very first case of Covid-19 till 14 of May 2020, therefore it is being witnessed that before pandemic the market was stable and have very least correlation but after the Coronavirus pandemic entire European capital market started to travel in the same direction highlighted as in red. As per the theory of Forbes and Rigobon (2002) the domino effect of one market to another one, if one market crashed in the same region then there many chances that it will affect some other market, and this what is seen in the correlation matrix after the very first case of Covid-19 to the current date, the correlation markets indicates market to market impact in American & European regions, and these are the regions which are badly affected by the virus.

Table 8 Regression Model for each index

| Regression Panel-1 | A | B | CI | DI | E | Y | Z | | | |
|---|---|---|---|---|---|---|---|---|---|---|
| Coefficient | -0.0153 | -0.015 | -0.014 | -0.016 | 0.066 | -0.0492 | -0.0140 | | | |
| t-Statistic | -2.008** | -2.095** | -2.014** | -1.828* | 1.549 | -3.3967 | -1.5033 | | | |
| Adjusted R-squared | 0.0361 | 0.040 | 0.036 | 0.028 | 0.017 | 0.1584 | 0.0232 | | | |
| Cases in Start | 1 | 1 | 1 | 1 | 1 | 1 | 5 | | | |
| Cases in End | 1,390,746 | 1,390,746 | 1,390,746 | 1,390,746 | 1,390,746 | 188,974 | 40,186 | | | |
| Regression Panel-2 | AG | AH | AI | AJ | AK | AL | AM | AN | AO | AP | AQ |
| Coefficient | -0.0049 | -0.0053 | -0.0057 | -0.0040 | -0.0031 | -0.0295 | -0.0218 | -0.0117 | -0.0122 | -0.0191 | -0.0041 |
| t-Statistic | -1.906** | -1.5506 | -2.129** | -1.4981 | -0.6476 | -3.574* | -3.098* | -1.4801 | -1.5109 | -2.433** | -0.8310 |
| Adjusted R-squared | 0.0272 | 0.0147 | 0.0363 | 0.0131 | -0.0072 | 0.1205 | 0.1396 | 0.0163 | 0.0175 | 0.0624 | -0.0057 |
| Cases in Start | 27 | 27 | 27 | 27 | 1 | 1 | 2 | 1 | 1 | 1 | 2 |
| Cases in End | 84,024 | 84,024 | 84,024 | 84,024 | 440 | 3,017 | 15,438 | 78,003 | 78,003 | 11,618 | 35,788 |
| Regression Panel-3 | G | H | I | J | K | L | M | N | O | P | Q |
| Coefficient | -0.009 | -0.021 | -0.019 | -0.001 | -0.015 | -0.001 | -0.008 | -0.027 | -0.013 | -0.0330 | -0.0206 |
| t-Statistic | -1.186 | -2.008** | -2.183** | -0.352 | -2.058** | -1.633* | -2.318** | -3.602* | -1.929** | -2.599*** | -2.925*** |
| Adjusted R-squared | 0.005 | 0.040 | 0.047 | -0.016 | 0.044 | 0.022 | 0.072 | 0.187 | 0.037 | 0.0932 | 0.0950 |
| Cases in Start | 1 | 2 | 3 | 1 | 1 | 3 | 1 | 2 | 1 | 2 | 1 |
| Cases in End | 172,239 | 229,705 | 140,734 | 43,211 | 228,691 | 222,104 | 30,330 | 28,132 | 53,981 | 15,964 | 27,909 |
| Regression Panel-3 | AA | AB | AC | AD | AE | R | X | | | | |
| Coefficient | -0.0160 | -0.0203 | -0.0083 | -0.0240 | -0.0097 | -0.0104 | -0.0191 | | | | |
| t-Statistic | -1.763* | -1.4585 | -2.514* | -2.6204 | -3.990*** | -1.944** | -2.403** | | | | |
| Adjusted R-squared | 0.0285 | 0.0154 | 0.0945 | 0.1050 | 0.2490 | 0.0482 | 0.0761 | | | | |
| Cases in Start | 2 | 2 | 1 | 2 | 1 | 1 | 2 | | | | |
| Cases in End | 242,271 | 242,271 | 17,204 | 3,380 | 143,114 | 10,667 | 16,548 | | | | |
| Regression Panel-3 | S | T | U | W | AF | | | | | | |
| Coefficient | -0.0120 | -0.0417 | -0.0037 | -0.0322 | -0.0102 | | | | | | |
| t-Statistic | -1.0107 | -2.813*** | -0.7262 | -1.720* | -2.063** | | | | | | |
| Adjusted R-squared | 0.0003 | 0.0824 | -0.0088 | 0.0245 | 0.0590 | | | | | | |
| Cases in Start | 1 | 4 | 1 | 3 | 1 | | | | | | |
| Cases in End | 16,079 | 6,975 | 1,147 | 25,346 | 44,830 | | | | | | |

\* p-value > 0.05 but < 0.10 or > 5% and < 10%

\*\* p-value > 0.01 but < 0.05 or > 1% and < 5%

\*\*\* p-value < 0.01 or < 1%

By using linear regression model, we quantified the relation of % change in Coronavirus cases to index returns, panel 1 consists on American indices, the indices from A to DI witnesses negative significant relationship between % change in cases to the index returns, as much higher the percentage as much will be a decline into the stock market, hence America is the highest affected region by Novel coronavirus and off course due to complete lockdown in the entire states of America brought significant decline into the capital market, many businesses closure and compressed of demand of basics products put the market in trouble, supporting to above statement these are the major indices in America which have potential represent entire American region collectively.

Panel 2 comprised on Asian region stock indexes, due to rapid increase in Covid-19 cases in China, Thailand and Philippine impacted heavily on the economy of these countries including the stock market as well, the outbreak took its first breath in Wuhan (China Mainland) and travels from China to the entire world, therefore table above showing the significant relationship with major indices of Chinese capital market, increment in Covid-19 cases reflects into Chinese capital market, even not China is only one which is lack behind, Thailand and Philippine also shown the impact of percentage change into Covid-19 cases with stock market downfall.

Panel 3 categorized one European based indexes, interestingly RTSI (Russia) and Budapest SE (Hungary) did not have any significant impact on the % change in Covid-19 cases with stock market return perhaps the had some smart policy to deal with the global pandemic, but rest of the entire European region capital markets are badly damaged by the virus pandemic that what was analyzed in the above-mentioned segment in the correlation matrix, we have further examined the effected indexes by assigning them ranks according to their Coefficients driven from linear regression model as below.

Table 9 Country wise Coefficients

| Indices Code | Country | Coefficients | Ranks |
|---|---|---|---|
| J | Netherland | -0.09% | 1 |
| L | Italy | -0.13% | 2 |
| M | Switzerland | -0.76% | 3 |
| AC | Poland | -0.83% | 4 |
| G | Germany | -0.90% | 5 |
| AE | Turkey | -0.97% | 6 |
| R | Denmark | -1.04% | 7 |
| O | Belgium | -1.31% | 8 |
| K | Spain | -1.49% | 9 |
| AA | Russia | -1.60% | 10 |
| X | Israel | -1.91% | 11 |
| I | France | -1.94% | 12 |
| Q | Sweden | -2.06% | 13 |
| H | UK | -2.11% | 14 |
| N | Portugal | -2.72% | 15 |
| P | Austria | -3.30% | 16 |

We have excluded Russian RTSI (Russia) and Budapest SE (Hungary) from the rankings because Covid-19 has no impact on these indexes, we used coefficient to account for the minor and major change in the

% change in cases bring the unit to change into index return, therefore above mentioned ranks allow us to pass comments that Austria is the highest influential country in term of coefficient increment, suppose 1% changes are detected in % change of cases so it will bring 3.30% change into the stock market returns of ATX index, although Worldsdometer ranks Italy, Spain, Germany, and Turkey into to ten effected countries but above grid shows indeed a significant inverse relationship between % change in cases and index return the effect is very nominal.

Panel 4 is quantified on Pacific and Gulf-based indexes, S&P/ASX 200, STI Index and Tadawul All Share show a significant negative relationship between coronavirus cases and stock market returns, the coefficient of the equations are very low that means there is a very minor type of effect on the indexes by the increment in the Covid-19 cases within the country, therefore the relationship still exists and can't be ignored at all.

Table 10 EGARCH Model for each index in Developed Market classified by MSCI

| EGARCH Model – Each Index for Developed Markets | | | | | | | |
|---|---|---|---|---|---|---|---|
| Code | Indices | MSCI | Coefficient | t-statistics | ARCH Term | Asymmetry term | GARCH Term |
| A | Dow 30 | Developed | 0.000 | 0.152 | 0.204** | -0.368*** | 0.958*** |
| B | S&P 500 | Developed | 0.001 | 0.001 | 0.283** | -0.394*** | 0.957*** |
| CI | Nasdaq | Developed | 0.002 | 1.067 | 0.270** | -0.316*** | 0.957*** |
| DI | SmallCap 2000 | Developed | -0.001 | -0.492 | 0.214** | -0.250*** | 0.978*** |
| E | S&P 500 VIX | Developed | -0.02 | -3.151*** | -0.066 | 0.413*** | 0.878*** |
| G | DAX | Developed | 0.006 | 3.814*** | -0.150** | -0.242*** | 0.965*** |
| H | FTSE 100 | Developed | 0.000 | 0.091 | 0.497*** | -0.138* | 0.948*** |
| I | CAC 40 | Developed | -0.006 | -3.529*** | 0.317*** | -0.151*** | 0.955*** |
| J | AEX | Developed | -0.003 | -1.042 | 0.689*** | -0.190* | 0.920*** |
| K | IBEX 35 | Developed | 0.003 | 4.433*** | -0.151*** | -0.283*** | 0.959*** |
| L | FTSE MIB | Developed | 0.006 | 6.165*** | -0.248*** | -0.369*** | 0.937*** |
| M | SMI | Developed | 0.005 | 4.109*** | 1.0317*** | -0.478*** | 0.119* |
| N | PSI 20 | Developed | -0.003 | -1.818* | 0.728*** | -0.10619 | 0.884*** |
| O | BEL 20 | Developed | 0.002 | 1.257 | 0.158*** | -0.277*** | 0.959*** |
| P | ATX | Developed | -0.002 | -0.600 | 0.156** | -0.185*** | 0.984*** |
| Q | OMXS30 | Developed | 0.006 | 3.609*** | -0.150** | -0.231*** | 0.962*** |
| R | OMXC25 | Developed | 0.002 | 1.358 | 0.232*** | -0.222*** | 0.897*** |
| S | Nikkei 225 | Developed | -0.000 | -0.39 | 0.066 | -0.216*** | 0.974*** |
| T | S&P/ASX 200 | Developed | -0.003 | -2.606** | 0.381*** | -0.081 | 0.957*** |
| U | DJ New Zealand | Developed | 0.001 | 0.605 | 0.169** | -0.127*** | 0.965*** |
| W | STI Index | Developed | -0.000 | -0.352 | 0.202** | -0.190*** | 0.958*** |
| X | TA 35 | Developed | -0.001 | -0.824 | 0.369*** | -0.123** | 0.975*** |

p-value > 0.05 but < 0.10 or > 5% and < 10%

** p-value > 0.01 but < 0.05 or > 1% and < 5%

*** p-value < 0.01 or < 1%

Not only we detected the effect and intensity of the Covid-19 on stock indices but also we have encountered clustering effects into each index classified as Developed and Emerging markets, to measure volatility in the stock indices we have used ordinary least square (OLS), GARCH, TARCH, PARCH and EGARCH models, as per the information criterions (AIC, HIC, and HQC) the least valuable one model is the best, we have selected to used EGARCH model in both of the panels, we have tested every single index to find the clustering effect in the markets by using dummy in place of reported cases on daily basis, below is the list of cumulative cases on the last terminal day of this research.

Table 11 Cumulative Cases in Developed Markets with respect to Country

| S.no | ISO3 Codes | Country | Market | Cumulative Cases | % of Cumulative Cases |
|---|---|---|---|---|---|
| 1 | USA | American | Developed | 1,390,746 | 52.27% |
| 2 | GBR | UK | Developed | 229,705 | 8.63% |
| 3 | ESP | Spain | Developed | 228,691 | 8.60% |
| 4 | ITA | Italy | Developed | 222,104 | 8.35% |
| 5 | DEU | Germany | Developed | 172,239 | 6.47% |
| 6 | FRA | France | Developed | 140,734 | 5.29% |
| 7 | BEL | Belgium | Developed | 53,981 | 2.03% |
| 8 | NLD | Netherland | Developed | 43,211 | 1.62% |
| 9 | CHE | Switzerland | Developed | 30,330 | 1.14% |
| 10 | PRT | Portugal | Developed | 28,132 | 1.06% |
| 11 | SWE | Sweden | Developed | 27,909 | 1.05% |
| 12 | SGP | Singapore | Developed | 25,346 | 0.95% |
| 13 | ISR | Israel | Developed | 16,548 | 0.62% |
| 14 | JPN | Japan | Developed | 16,079 | 0.60% |
| 15 | AUT | Austria | Developed | 15,964 | 0.60% |
| 16 | DNK | Denmark | Developed | 10,667 | 0.40% |
| 17 | AUS | Australia | Developed | 6,975 | 0.26% |
| 18 | NZL | New Zealand | Developed | 1,147 | 0.04% |
| Total | Total | | | 2,660,508 | 100.00% |

It has discussed previously, America found the highest affect country around the globe; around 1.3m cases are reported in US and New Zealand as the least cases around the world. Anyhow, according to above mentioned EGARCH equation witnessed the smart policies of Portugal and Australia because the PSI 20 & S&P/ASX 200 asymmetry term is insignificant, which indicates that in both of the indexes there is no instable fluctuation which harms market decorum, therefore it is also noticeable that these both market are saved from markets shocks generated by the bad news. Coming towards the rest of the indices, almost every index reported the clustering effects because p-value of the EGARCH Term is under accepted regions, meaning we cannot reject the alternative hypothesis, there are clustering effect in the model, meaning period of low volatility is followed by period of low volatility for prolonged period and period of high volatility is followed by period of high volatility, if one day return is negative then there possibilities that next will be negative too and this pattern remains same with a certain time, and what has been seen in the rest of the indexes in developed market. It is also noticeable that most of the indexes categorized into the developed market are from America or Europe and that is the red zone area of Covid-19, since it is declared a global pandemic entire markets shows negatives returns mentioned by the model above.

Table 12 EGARCH Model for each index in Emerging Market classified by MSCI

| | | | EGARCH Model – Each Index for Developed Markets | | | | |
|---|---|---|---|---|---|---|---|
| Code | Indices | MSCI | Coefficient | t-statistics | ARCH Term | Asymmetry term | GARCH Term |
| Y | Bovespa | Emerging | -0.002 | -0.756 | 0.254*** | -0.255*** | 0.945*** |
| Z | S&P/BMV IPC | Emerging | 0.001 | 2.410*** | -0.142*** | -0.250*** | 0.958*** |
| AA | MOEX | Emerging | 0.000 | 0.146 | 0.201*** | -0.128*** | 0.967*** |
| AB | RTSI | Emerging | 0.000 | 0.227 | 0.179 | -0.140*** | 0.971*** |
| AC | WIG20 | Emerging | 0.003 | 2.368** | -0.122*** | -0.250*** | 0.962*** |
| AD | Budapest SE | Emerging | -0.001 | -0.277 | 0.178** | -0.127*** | 0.970*** |
| AE | BIST 100 | Emerging | 0.001 | 0.938 | 0.000 | -0.239*** | 0.930*** |
| AF | Tadawul All Share | Emerging | 0.007 | 3.824*** | 0.072 | -0.265*** | 0.896*** |
| AG | Shanghai | Emerging | -0.001 | -1.551 | 1.076*** | -0.136* | -0.158 |
| AH | SZSE Component | Emerging | 0.003 | 1.976** | 0.111 | -0.226*** | 0.889*** |
| AI | China A50 | Emerging | 0.000 | 0.948 | 0.037 | -0.264*** | 0.906*** |
| AJ | DJ Shanghai | Emerging | -0.002 | -1.339 | 0.797 | -0.149 | -0.205 |
| AK | Taiwan Weighted | Emerging | 0.005 | 3.324 | 0.034 | -0.207*** | 0.935*** |
| AL | SET | Emerging | 0.002 | 2.921** | -0.146*** | -0.298*** | 0.961*** |
| AM | IDX Composite | Emerging | -0.001 | -0.443 | 0.107 | -0.197*** | 0.965*** |
| AN | Nifty 50 | Emerging | 0.002 | 2.168** | 0.041 | -0.226*** | 0.980*** |
| AO | BSE Sensex | Emerging | 0.002 | 2.144** | 0.022 | -0.233 | 0.98 |
| AP | PSEi Composite | Emerging | -0.012 | -9.487*** | 1.138*** | 0.057 | -0.203*** |
| AQ | Karachi 100 | Emerging | 0.000 | 0.277 | 0.168** | -0.106** | 0.942*** |

\* p-value > 0.05 but < 0.10 or > 5% and < 10%

\*\* p-value > 0.01 but < 0.05 or > 1% and < 5%

\*\*\* p-value < 0.01 or < 1%

In the emerging market, most of the countries are from Asian and Pacific regions, apart from the China cross border countries to China are declared red zone in the pandemic. In above table Fourth and Fifth column is the mean equation in the EGARCH model and rest of the columns for the EGARCH, result from the emerging market are far different from the developed market such as the biggest index of china and India DJ Shanghai and BSE Sensex had no clustering effect on pandemic even PSEi Composite does not have any clustering effect because p-value of Asymmetry term (EGARCH Model) more than 0.10% which means negative news or shock does not impact the market equilibrium, therefore smart policies and initiative to retain the capital taken by the Indian, Chinese and Philippine looks fruitful stabilized the market equilibrium. EGARCH refers to two broad criteria such as it measures the volatility into the stock market as well as the role of information meaning negative and positive shocks into the market, before moving forward we mentioned the table which indicated cumulative cases in Emerging Markets.

Table 13 Cumulative Cases in Emerging Markets with respect to Country

| S.no | ISO3 Codes | Country | Market | Cumulative Cases | % of Cumulative Cases |
|---|---|---|---|---|---|
| 1 | RUS | Russia | Emerging | 242,271 | 26.67% |
| 2 | BRA | Brazil | Emerging | 188,974 | 20.81% |
| 3 | TUR | Turkey | Emerging | 143,114 | 15.76% |
| 4 | CHN | China | Emerging | 84,024 | 9.25% |
| 5 | IND | India | Emerging | 78,003 | 8.59% |
| 6 | SAU | Saudi Arab | Emerging | 44,830 | 4.94% |
| 7 | MEX | Mexico | Emerging | 40,186 | 4.42% |
| 8 | PAK | Pakistan | Emerging | 35,788 | 3.94% |
| 9 | POL | Poland | Emerging | 17,204 | 1.89% |
| 10 | IDN | Indonesia | Emerging | 15,438 | 1.70% |
| 11 | PHL | Philippine | Emerging | 11,618 | 1.28% |
| 12 | HUN | Hungary | Emerging | 3,380 | 0.37% |
| 13 | THA | Thailand | Emerging | 3,017 | 0.33% |
| 14 | TAI | Taiwan | Emerging | 440 | 0.05% |
| Total | Total | | | 908,287 | 100.00% |

Russia, Brazil, and Turkey show the highest reported cases the in the above-mentioned table, whereas Chine remains on the number, therefore in the indexes of Russia, Brazil and Turkey strong clustering effect has been detected by EGARCH model, whereas in rest of the Chinese indices show aggressive clustering effect apart from Shanghai index, we have dummy variable in the palace of every reported case to developed EGARCH equation, hence almost every index individually effect by the Covid-19 reported by the regression model above and there are few indexes which have Covid-19 impact but not clustering effect. Further, in the above-mentioned table, the emerging market indicates that there is a clustering effect in each index apart from the Shanghai, BSE Sensex, and PSEi Composite.

Table 14 Market wise Classification of EGARCH Model

| | Developed – Emerging Market | | | | | | |
|---|---|---|---|---|---|---|---|
| | EGARCH MODEL | | | | | | |
| | Developed Market | | | | Emerging Market | | |
| | Coeff. | z-stat. | Prob. | | Coeff. | z-stat. | Prob. |
| Mean Equation | | | | Mean Equation | | | |
| C | 0.00032 | 1.025625 | 0.3051 | C | 0.000154 | 0.350885 | 0.7257 |
| Developed Market | 0.002213 | 3.31E+00 | 0.0009 | Emerging Markets | 0.001389 | 2.14E+00 | 0.0321 |
| AR (1) | 0.177919 | 2.669502 | 0.0082 | AR (1) | 0.187777 | 2.822439 | 0.0052 |
| Variance Equation | | | | Variance Equation | | | |
| C | -0.26665 | -3.06681 | 0.002 | C | -0.82169 | -3.43228 | 0.0006 |
| ARCH Term | -0.03495 | -0.6348 | 0.526 | ARCH Term | 0.393685 | 4.363591 | 0.0000 |
| Asymmetry term | -0.26024 | -7.62018 | 0.000 | Asymmetry term | -0.14025 | -2.61873 | 0.0088 |
| GARCH Term | 0.972621 | 138.7587 | 0.000 | GARCH Term | 0.948585 | 42.9176 | 0.0000 |
| ARCH LM Test | | | | ARCH LM Test | | | |
| | | t-Statistic | Prob. | | | t-Statistic | Prob. |
| | | 0.108679 | 0.9136 | | | -0.23796 | 0.8121 |

In every individual phase of this research paper, we found European market is the most affected market by the Covid-19 and most of the indexes are in the developed market comes under European region, therefore we needed to hypothesize the Covid-19 clustering effect on Developed and Emerging Market collectively, before this segment we analysis the market on an individual basis by using regression and EGARCH Model, but in this segment, we have averaged out the daily return of each index and plugged this into the respective category and created single indexes for developed and emerging markets.

Initially, we tested the ARCH effect in the both constructed equation and found ARCH effect because p-value of AR(1) is less than 0.05 or 5% which allow us to use ARCH family model, therefore we use Ordinary Least Square, GARCH, TARCH, EGARCH and PARCH model, we selected EGARCH model according to least value of AIC, SIC, and HQC.

Interesting both of the Developed and Emerging Markets exhibited positive returns in these 11 months on the average basis shown by the mean driven equation the first part of the model, invariance equation, it has been witnessed that developed and emerging markets have strong clustering effect because p-value of Asymmetry term is negative and less than 0.05 or 5% which further claims that the negative shock within the market effect more rather than the positive shocks, which means huge increment in cumulative cases make significant persistent volatility for long period, further coefficients of the mean and variance equation are higher in the Developed Market compare to the Emerging Market, meaning Developed Market is more volatile and have persistent clustering effect. Finally, we use diagnostic as ARCH-LM test which indicates that equations do not have the ARCH type of effect; therefore p-value of ARCH-LM test in both markets is more than 0.05 or 5% which proclaims that both equations do not have the ARCH type of effect.

Table 15 Continent wise Classification of EGARCH Model

| Continent Classification | | | | | | | |
|---|---|---|---|---|---|---|---|
| EGARCH MODEL | | | | | | | |
| | American Market | | | | Asian Market | | |
| | Coeff. | z-stat. | Prob. | | Coeff. | z-stat. | Prob. |
| Mean Equation | | | | Mean Equation | | | |
| C | 0.00095 | 0.480074 | 0.6312 | C | -0.000181 | -0.405055 | 0.6854 |
| American Market | 0.0000438 | 1.91E-02 | 0.9847 | Asian Market | -0.004955 | -1.92E+00 | 0.0549 |
| AR (1) | 0.303847 | 4.70879 | 0.000 | AR (1) | 0.201858 | 3.043132 | 0.0026 |
| Variance Equation | | | | Variance Equation | | | |
| C | -8.618946 | -0.54581 | 0.585 | C | -0.521984 | -3.676844 | 0.0002 |
| ARCH Term | 0.01 | 0.182267 | 0.855 | ARCH Term | 0.200308 | 3.192093 | 0.0014 |
| Asymmetry term | 0.01 | 0.357303 | 0.721 | Asymmetry term | -0.150435 | -4.340833 | 0.000 |
| GARCH Term | 0.01 | 0.00551 | 0.996 | GARCH Term | 0.961391 | 78.37841 | 0.000 |
| ARCH LM Test | | | | ARCH LM Test | | | |
| | | t-Statistic | Prob. | | | t-Statistic | Prob. |
| | | 4.604761 | 0.000 | | | -0.938532 | 0.349 |
| | European Market | | | | Pacific & Gulf Market | | |
| | Coeff. | z-stat. | Prob. | | Coeff. | z-stat. | Prob. |
| Mean Equation | | | | Mean Equation | | | |
| C | 0.000279 | 0.522072 | 0.6016 | C | -0.0000134 | -0.031752 | 0.9747 |
| Europeon Market | -0.097015 | -9.49E-01 | 0.3428 | Pacific & Gulf Market | 0.000786 | 8.76E-01 | 0.3809 |
| AR (1) | 0.268005 | 4.098165 | 0.0001 | AR (1) | 0.390356 | 6.260474 | 0.000 |
| Variance Equation | | | | Variance Equation | | | |
| C | -0.35313 | -3.264434 | 0.001 | C | -0.464996 | -3.427611 | 0.0006 |
| ARCH Term | 0.142783 | 2.415525 | 0.016 | ARCH Term | 0.138435 | 1.815372 | 0.0695 |
| Asymmetry term | -0.2156 | -6.277141 | 0.000 | Asymmetry term | -0.2078 | -5.46053 | 0.000 |
| GARCH Term | 0.973524 | 99.29631 | 0.000 | GARCH Term | 0.964017 | 84.55819 | 0.000 |
| ARCH LM Test | | | | ARCH LM Test | | | |
| | | t-Statistic | Prob. | | | t-Statistic | Prob. |
| | | 0.096569 | 0.9232 | | | 1.335432 | 0.1831 |

As per the Morgan Stanley Capital International region classification guideline, we have further divided our model continent wise; the above table has the segments, (1) Mean Equation and (2) Variance Equation. We can understand the Covid-19 cases globally by the below-mentioned table and pie chart.

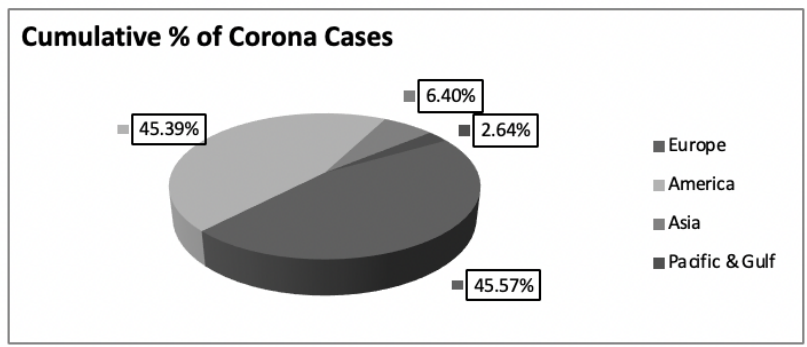

| Continents | Cumulative Cases | % |
|---|---|---|
| Europe | 1,626,184 | 45.57% |
| America | 1,619,906 | 45.39% |
| Asia | 228,328 | 6.40% |
| Pacific & Gulf | 94,377 | 2.64% |
| Total | 3,568,795 | 100.00% |

The table above shows the cumulative cases and percentage of the Covid-19 cases on the last terminal day of this research as of May 15, 2020. Europe stood on rank 1 by having around 45.57% cases in the continental ranking, while testing the Covid-19 Effect in correlation and regression model we found the heavy intervention of pandemic to shock the European markets, but in this segment, we have to test the volatility or clustering effect in each of the continents, we have averaged out the returns of all indices which come under European region and us single index as European index then we regressed this on the cumulative cases in the European region, in mean equation AR(1) show p-value less than 0.05 or 5% which allow us to use the ARCH family model, as per the Information criterion AIC, SIC, and HQC we selected EGARCH model is the most suitable model. p-value of Asymmetry term in European Market Model is less than 0.05 or 5% with a negative coefficient, which supports the negative shocks impact more rather than the positive news, Covid-19 is one of the kinds of a negative shock to the stock market and the value of the coefficient of European Market Model the highest amongst all of them for models, that further indicates the presence of stock market anomalies with market clustering in European indices, moreover below is the bar chart of countries in the European region with cumulative percentages Covid-19 Cases on last terminal day of this research.

By using the same methodology as above, we have averaged out all index and constructed the single index as Pacific and Gulf, hence again AR(1) do not limit us to use ARCH family model, therefore by using EGARCH Model we have driven mean and variance equation and came to the findings by defining the clustering effect into the region, in the last segment we employed diagnostic test as ARCH-LM test and the p-value is more than 0.05 or 5% which means there is no ARCH type of effect in the model and model is fit to predict results. The further below-mentioned graph illustrates country name along with % of cumulative Covid-19 cases in the European region.

*See Figure 3 uploaded in figure section.*

By using same methodology as above, in pacific region, which is considered the second least affected continent by the Covid-19 but it has the second-highest clustering effect, because the countries under this region started lockdown earlier then Asia and the region has more reported Covid-19 cases compare to Asia, which became the cause of clustering effect into the Pacific and Gulf region, however market exhibits the instability in term of daily volatility witnessed by the negative coefficient and p-value (< 0.05 or < 5%) of Asymmetry term, initially ARCH test did not limit us to use ARCH family models, hence we employed GARCH, TARCH, EGARCH and PARCH model, and found EGARCH as the most suitable model to measure Covid-19 clustering effects into Pacific and Gulf Market which further proclaims that negative news or shocks impact a lot than the positive news of shocks to the stock market, therefore rapid increment in Covid-19 on daily basis created significant negative news/shocks and that is noticeable sign for the clustering effect into the Pacific and Gulf region. In the last segment, we employed a diagnostic test as ARCH-LM test and the p-value is more than 0.05 or 5% which means there is no ARCH type of effect in the model and the model is fit to predict results. Further, the graph below exhibits the comprised countries and % cumulative Covid-19 cases on the last terminal day of this research.

*See Figure 4 uploaded in figure section.*

To drive result for Asian Stock market we use the same pattern of calculating index and regressed it for EGARCH model as above, initially we applied the ARCH test to see whether long term volatility is followed by another period of long term volatility or in easy words can ARCH family models be used to witness clustering effect into the region, hence the test allowed us to use ARCH family models, we found EGARCH model is the best-suited model, Asian markets reported the negative abnormal returns, meaning as Covid-19 cases increases in the market produce the negative downwards, going forward it has also been noticed that market has strong clustering effect in the pandemic period. According to worlddometer and rest of the other authentic sources, Asia is the least affected region by Covid-19 (Apart from China), two factors became the cause of cushion or savior for Market from the virus effect, (1) Smart LockDown in the Market and (2) Rapid Recoveries from the Virus, further, we can't ignore the role of cumulative cases in Asia which still lesser than rest of the other regions, and we found the least volatility into the market. We employed the ARCH-LM test to check model fitness and the p-value is less than more than 0.05 or 5% which further indicates there is no ARCH type of effect in the model and model is fit to use for prediction. Below is the graph which illustrates % of Covid-19 cases in the pandemic.

*See Figure 5 uploaded in figure section.*

America is considered second the highest affected region by Covid-19, interestingly we could not find clustering effect into American stock market, by using the same methodology as above we employed the EGARCH model and found indeed there is a significant relationship between individual American Index and Covid-19 cases, but there is no clustering effect as illustrated by the value of the Asymmetry term market is not effect by the negative shock or news in the pandemic, in the light of the above numbers we can claim that America uses good Standard Operating Procedure (SOPs) to stabilize the capital market in America. ARCH-LM test defines the model fitness because p-value is more than 0.05 or 5%, further below is the graph of % of Covid-19 cases in the American Countries.

*See Figure 6 uploaded in figure section.*

We have also constructed model with continent indexed daily average returns with cumulative cases in the continent and also with dummy variable by replacing cumulative cases into EGARCH model, employed data is divided into categories (1) before the Covid-19 cases not a single case reported in the region (we use 0 as dummy variable) and (2) After the very first cases Covid-19 (we use 1 as dummy variable). We found the first model not suitable because finding were against logics so we dropped that model and decided to select the second model with the dummy variable and the model is with the logic and realities.

# 5. Discussions And Conclusion

**5.1 Discussion:** the purpose of this study is to encounter Covid-19 impact on global indices; hence we have selected the best performing indices around the globe classified by the investing.com, we have used umbrella approach to see market movement before and after pandemic by using different models, in the first phase of the research we used descriptive statistics by classifying it region-wise as per the methodology of MSCI, in which we see the median shift, standard deviation shift and relative rank shifts before and after the pandemic, plus also try to relate the cumulative cases with these shifts into each sampled index. In the second phase, we have seen the correlation matrix where is used to see the joint movement of entire sampled indexes to each other before and after the pandemic, in the third phase we tested linear regression model for each index by categorizing them region-wise as per the MCSI classification to see the impact of % change into Covid-19 with single individual index, not only the impact of Covid-19 to index return we also encountered the clustering effect-Volatility in each index by using EGARCH model in this is classified as Developed and Emerging Market as per the methodology of MSCI this is in phase four. With the same method by dividing markets into two broad segments as Developed and Emerging Market, we collectively tested clustering effect by using EGARCH model in phase five, in the last phase six we classified the indexes into four broad categories as America, Asia, Europe, and Pacific & Gulf by assigning them a single index and constructed variance equation by using EGARCH model.

**5.2 Conclusion:** as per international reports, articles and other authentic source, region Europe found the highest affected region around the globe, hence our research found the European Market in much vulnerable and risk, shifts in descriptive statics shown the trend how European market goes from better to good and worse finally, in these 11 months shifts indicates not only returns are affected by the Covid-19 in European capital market but also ranking with peer group became worsen, further the correlation matrix illustrated after the rapid increase in the Novel Coronavirus Cases entire European market exhibited highly correlated market, means almost all big indexes in Europe move closely to each other and the linear regression model supported the this movement by illustrating the significant impact on the Market Returns, moreover in MSCI index classification most indexes in Developed Market are from European countries, due to this Developed ARCH model reported clustering effect in the developed markets, and finally in the last model we have found European Market has the highest volatility in relation to Covid-19.

With the rapid increase of Covid-19 cases in America affected American indexes, we found that American indexes exhibited significant shifts in term of average monthly returns, standard deviation, and relative rankings before and after the announcement of the pandemic by World Health Organization, we found American Indices as stable because of intellectual and smart investment policies for capital markets, indeed there is a significant and strong impact of Covid-19 on indices returns but no evidence found for clustering effect/volatility in American indices.

Pacific & Gulf region countries reported the third-highest Covid-19 cases, therefore we found the great shift in medians, standard deviation and relatives ranking in Pacific & Gulf indices, further we also detected Covid-19 impacts on indexes daily returns, therefore in the last segment we found the Pacific & Gulf indexes are classified as the second highest volatile market having strong clustering effect in contrast to Covid-19 cases.

Due to fewer cases reported in the Asian region and smart lockdown policy, we found that Asian stock index indeed has Covid-19 impact on indexes but least clustering effect compare to Europe and Pacific & Gulf, median shift indicated that after the declaration of pandemic (11-March-2020) market reported a significant change in each index on Asian capital Market, therefore we can conclude Asian Capital Market has the third effect market by the Covid-19.

# Declarations

It has been declared that there is no conflict of interest any of the authors

# Figures

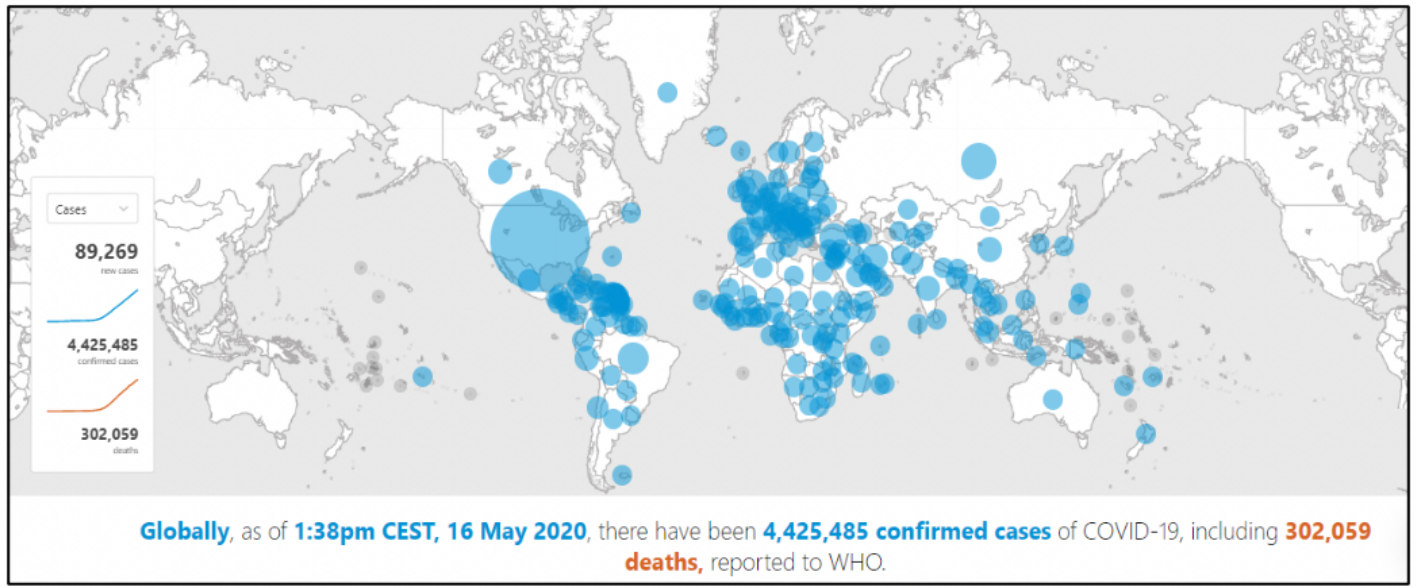

### Figure 1

WHO Coronavirus Disease (COVID-19) Dashboard. Note: The Graph has taken from the WHO official website on 16th May 2020, the graph consists of the number of new cases, the aggregate number of cases, and the number of death on the left-hand side, size of the circle showing the number of cases in the region or country. Note: The designations employed and the presentation of the material on this map do not imply the expression of any opinion whatsoever on the part of Research Square concerning the legal status of any country, territory, city or area or of its authorities, or concerning the delimitation of its frontiers or boundaries. This map has been provided by the authors.

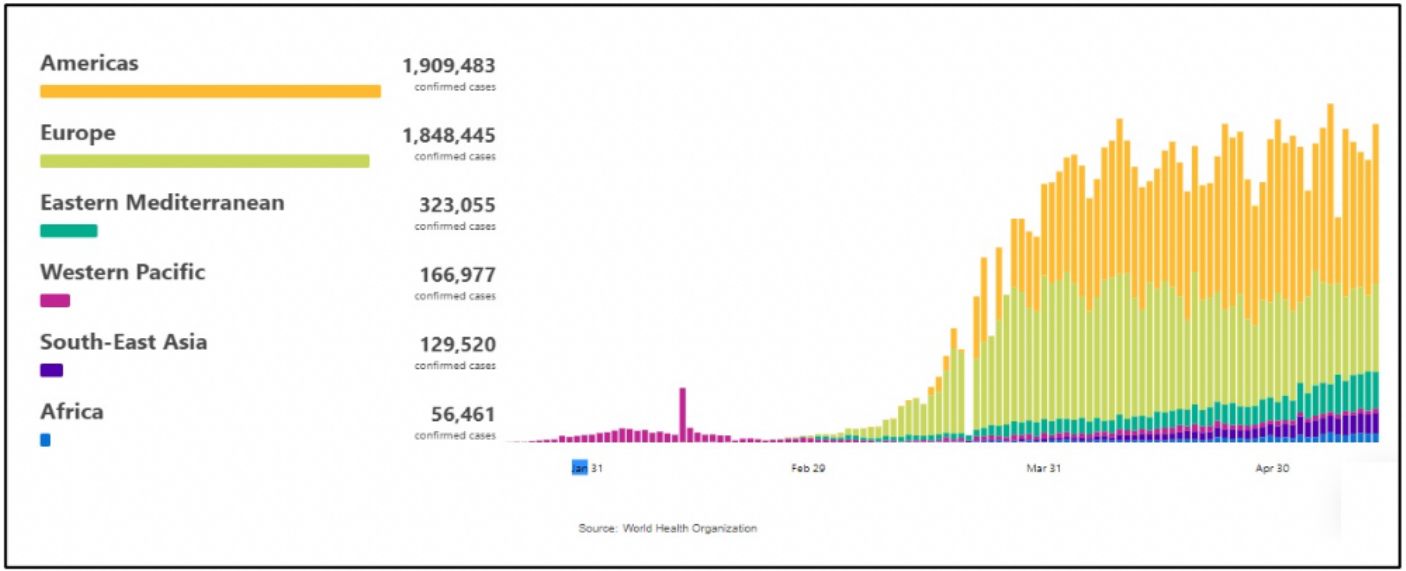

Figure 2

Case Comparison WHO Regions. Note: The graph is taken from the official website of WHO on 16th May 2020, which represented the number of cases according to the subcontinent, the graph indicating that most cases have been reported in the America and Europe subcontinent, further if classified the market most of the developed market belongs to America and Europe Countries.

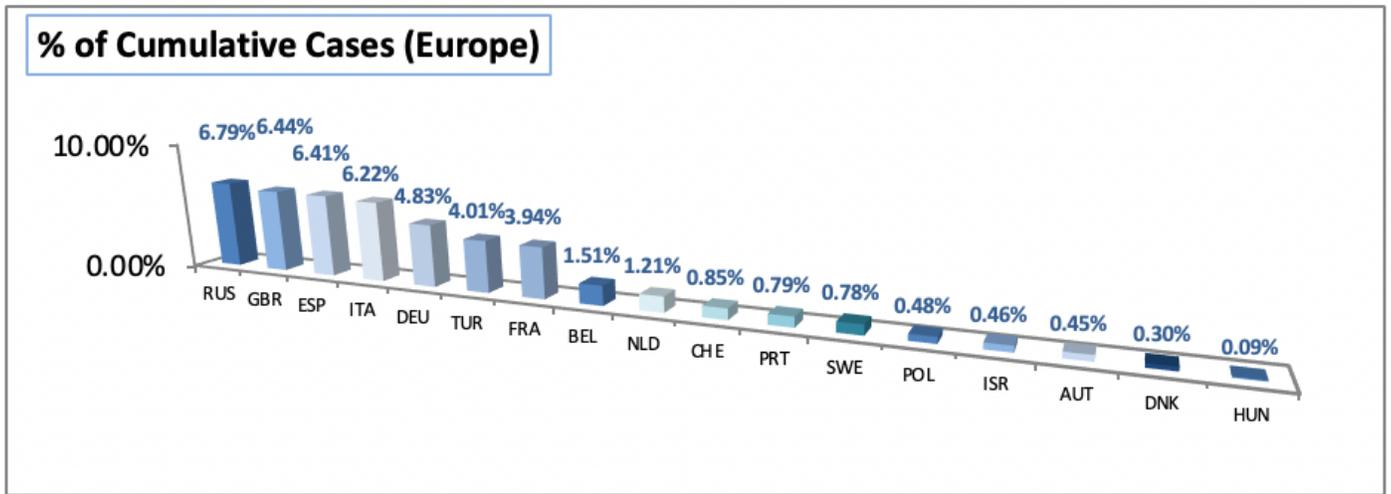

Figure 3

European region

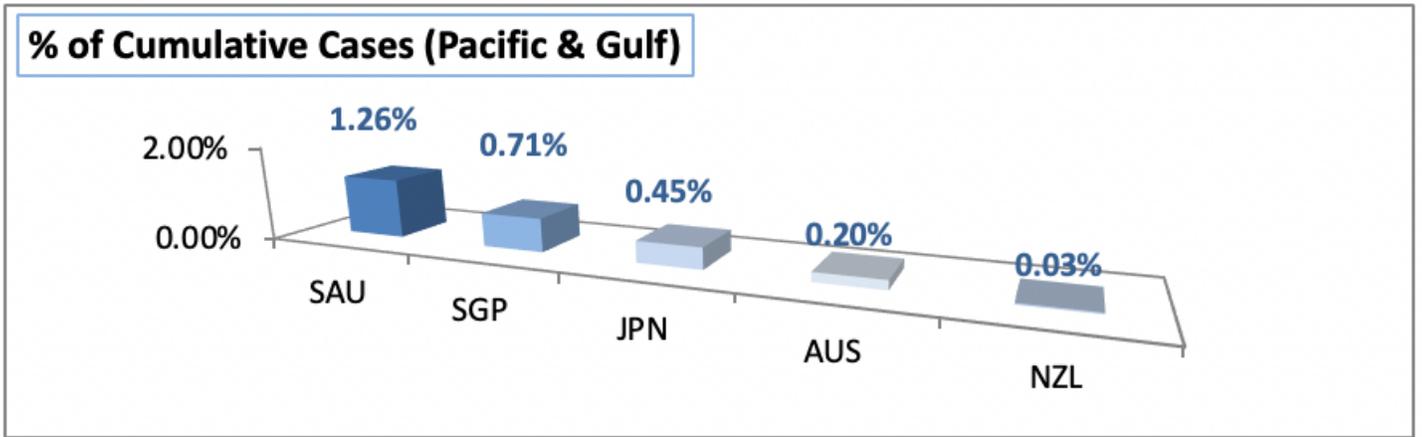

Figure 4

Pacific and Gulf region

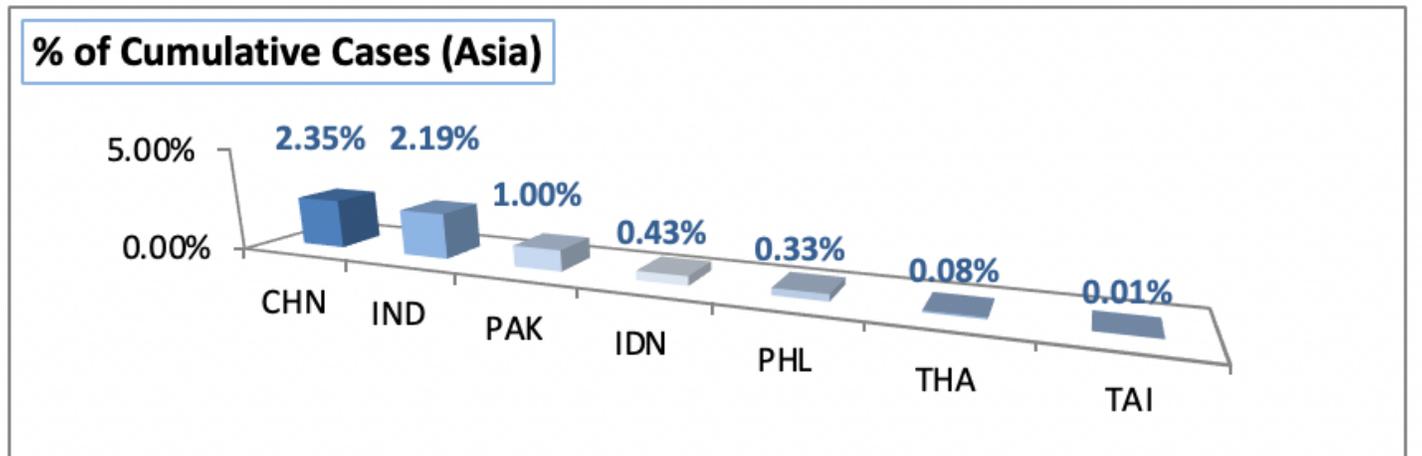

Figure 5

Asia

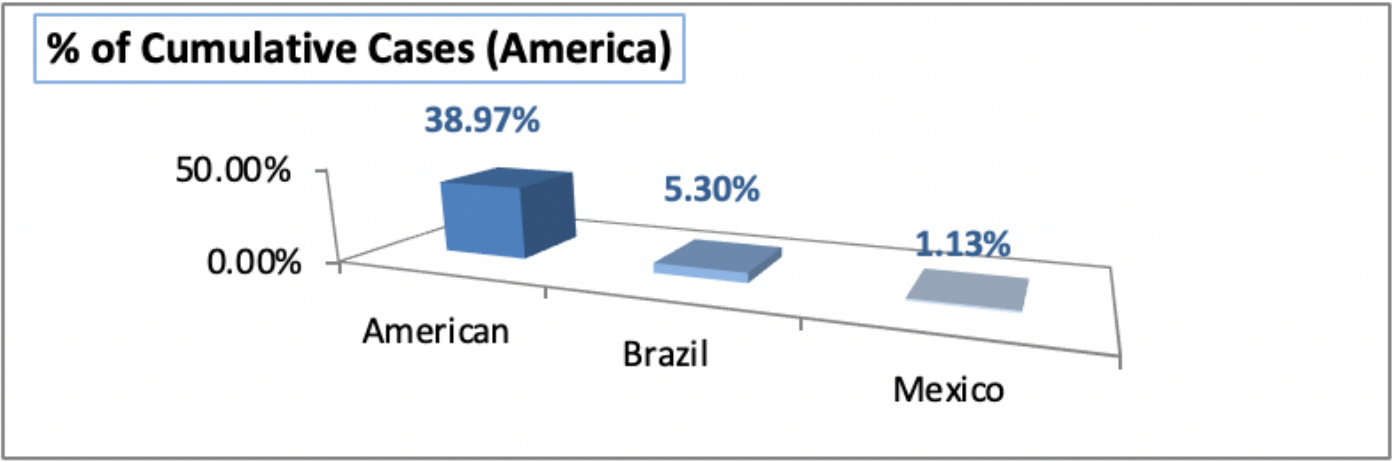

Figure 6

America